\documentclass[useAMS,usenatbib]{mn2e}

\usepackage{graphicx,color,afterpage}
\usepackage[total={17.8cm,24.0cm},centering]{geometry}
\usepackage{times}

\def\hi{H{\small I}}

\def\kms{km~s$^{-1}$}

\def\arcsec{$^{\prime \prime}$}
\definecolor{Mygrey}{gray}{0.75}

\newcommand{\ltsimeq}{\raisebox{-0.6ex}{$\,\stackrel{\raisebox{-.2ex}{$\textstyle <$}}{\sim}\,$}}
\newcommand{\gtsimeq}{\raisebox{-0.6ex}{$\,\stackrel{\raisebox{-.2ex}{$\textstyle >$}}{\sim}\,$}}
\newcommand{\farc}{\mbox{\ensuremath{.\!\!^{\prime\prime}}}}

\usepackage[compact]{titlesec}
\titlespacing{\section}{0pt}{*2}{*1}

\title[Molecular gas in LIRGs; Arp~157]{The molecular ISM in luminous infrared galaxies: a $\lambda$=3mm line survey of Arp~157} 
\author[Timothy A. Davis et al.]{Timothy A. Davis$^{1}$\thanks{E-mail: \texttt{tdavis@eso.org}}, Amanda Heiderman$^{2}$, Neal J. Evans II$^{3}$ and Daisuke Iono$^{4}$
 \vspace{0.4cm}\\
\parbox{\textwidth}{$^{1}$European Southern Observatory, Karl-Schwarzschild-Str. 2, 85748, Garching-bei-M\"unchen, Germany\\
$^{2}$Max-Planck-Institut f\"ur Astronomie, K\"onigstuhl 17, D-69117 Heidelberg, Germany\\
$^{3}$Department of Astronomy, University of Texas at Austin, 1 University Station C1400, Austin, TX 78712-0259, USA\\
$^{4}$National Astronomical Observatory of Japan, 2-21-1 Osawa, Mitaka, Tokyo 181-8588\\
}}

\begin{document}

\date{Accepted 2013 August 21.  Received 2013 August 20; in original form 2013 June 7}

\pagerange{\pageref{firstpage}--\pageref{lastpage}} \pubyear{2012}

\maketitle

\label{firstpage}

\begin{abstract}
 In this work we present a search for molecular line emission in the interacting luminous infrared galaxy Arp~157 (NGC 520). We report the detection of 19 different lines, from 12 different molecular species in our $\lambda$=3mm line survey and supporting $\lambda$=2mm and 1mm observations.
 We investigated the profiles of the detected lines, and find all species show asymmetric emission lines, with profiles that differ systematically as a function of critical density. 
We derive the physical conditions within the gas reservoir of this system using isotope ratios, an excitation analysis, and rotation diagrams. We suggest that in this source we are observing a warm (T$_{\rm ex}\gtsimeq$23K) edge-on, asymmetric starburst ring, plus a less dense central molecular component, which is more optically thin.
We compare our molecular abundances with several well studied galaxies, and with Galactic regions, in order to determine the primary driver of the chemistry in Arp~157. 
We highlight some significant differences between the abundances of molecules in this source and in the more extreme ultra-luminous infrared galaxy Arp 220. This suggests that different physical mechanisms may become important in driving gas chemistry at different points in the evolution of a merger induced starburst.
Of the nearby galaxies we considered, NGC~253 (whose nuclear chemistry is dominated by large-scale shocks) and NGC 1068 (whose nuclear chemistry is X-ray dominated) have the most similar molecular abundances to Arp~157. The Galactic region Sgr~B2(OH) (where shock desorption of ices is thought to be important) may also have similar abundances to Arp~157.
We hence postulate that shocks and/or the generation of hard radiation fields by the ongoing merger (or a starburst wind), are driving the molecular chemistry in Arp~157.
\end{abstract}


\section{Introduction}

The majority of the stellar mass of our universe was formed in high redshift galaxies, which host strong starbursts in dense, highly turbulent discs \citep[e.g.][]{2009ApJ...706.1364F}. 
The conditions in these systems are much more extreme than those found in normal star-forming galaxies in the local universe. 
Nearby merging galaxies with strong starbursts are the only places in the $z$=0 universe where we can study star formation in conditions that mimic those under which the majority of stars formed. Understanding local merger driven starbursts is thus vital to constrain theories of galaxy assembly and evolution.

The physical conditions within the interstellar-medium of a starburst may be quite different to those found in relatively quiescent local spiral and irregular galaxies. Extreme systems may have a different X$_{\rm CO}$ factor \citep[e.g.][]{2012ApJ...746...69G}, larger diffuse molecular envelopes than normal star-forming galaxies, and/or high dense gas fractions. The chemical composition of starburst galaxies is also different, and depends strongly on the main mechanisms inputing energy into the dense gas. Strong UV radiation fields, X-ray sources, shocks and cosmic rays all deposit energy into the ISM in different ways, altering its molecular chemistry \citep[e.g.][]{2009ApJ...696.1466B,2010A&A...519A...2G,2012ApJ...756..157G}.  Which of these heating mechanisms is most important in controlling the chemistry of merging luminous infrared galaxies (LIRGs) has yet to be established.

In this work, we aim to determine the physical conditions within the molecular gas reservoir of interacting LIRG Arp~157 (also known as NGC520), and investigate the effect of the ongoing merger on molecular abundances. 
To that end, we present an unbiased search for molecular emission in this galaxy at $\lambda$=3mm (and supporting data from the $\lambda$=2mm and 1mm atmospheric windows). 
The data were obtained when additional time became available whilst observing an {HCN} survey of the \textbf{V}IRUS-P \textbf{I}nvestigation of the e\textbf{X}treme \textbf{EN}vironments of \textbf{S}tarbursts; (VIXENS; \citealt{2011nha..confE..29H}) galaxies. HCN observations of the other VIXENS galaxies will be presented in a future work (Davis et al., in prep).

Arp~157 is one of the brightest disturbed galaxies in the sky \citep{1987QB857.A76......}. It has two small tidal tails, two nuclei and (at least) two
velocity systems in its spectra (see Section \ref{linesfound}), indicative of a late stage merger which has yet to relax. It is as
radio and infrared bright as the Antennae galaxies.

Numerical simulations \citep{1991ApJ...370..118S} indicate
that Arp~157 contains two interacting disks which collided
$\approx3\times10^{8}$ years ago.
This system has a nuclear starburst ring around the south-east nucleus (with an angular size of $\simeq6$\arcsec)
which can be resolved into $\sim10-15$ clumps in high resolution radio continuum imaging (which may correspond to supernova remnants; \citealt{2003MNRAS.346..424B}).
CO observations suggest $\approx1.9\times10^{9}$ M$_{\odot}$ of molecular gas is
concentrated in an east-west ring-like structure
coincident with these radio sources \citep{2001ApJ...550..104Y}. 
No molecular gas is seen near the
NW nucleus (which is separated from the SE nucleus by $\approx$40\arcsec), suggesting this progenitor disk was
rather gas-poor.
Optical spectra of the regions outside of this SE starburst nucleus show strong A star features, suggesting star-formation was more widespread  as the initial merger took place \citep{1991ApJ...381..409S}.
The starburst nucleus produces stars at a rate of $\approx18 M_{\odot}$\,yr$^{-1}$ (derived from narrow band Br$\gamma$ imaging), in a very compact area, and is the current
dominant source of star formation in this system \citep{2001A&A...366..439K}. Infrared estimates of the star formation rate are similar, e.g 13.9$\pm$0.8 M$_{\odot}$ yr$^{-1}$ based on IRAS data \citep{2011nha..confE..29H}. We observed our molecular line survey towards the SE nucleus, centred on coordinates RA=01:24:34.9, Dec=+03:47:30.1.

In Section \ref{data} we describe our observations with the IRAM-30m telescope, and our reduction strategy, and present the data. In Section \ref{results} we present our results. We first comment on the species detected and their observed line profiles in Section \ref{linesfound}. We then go on to discuss the physical conditions within the gas reservoir of Arp~157, derived from excitation analyses (Section \ref{cotex}), and from rotation diagrams (Section \ref{rotdiag_sec}). In Section \ref{chem_abund} we discuss the abundances of the detected molecular tracers, comparing to other galaxies (Section \ref{chem_abund_gals}), and selected Galactic regions (Section \ref{chem_abund_regions}), before concluding in Section \ref{conclude}.  A distance to the observed nucleus of Arp~157 of 28\,Mpc, and a recession velocity of 2258 \kms\ (which was confirmed with our molecular data) are assumed in this paper \citep{1988ngc..book.....T}, hence 1\arcsec\ corresponds to $\simeq$133\,pc.

\section{Data}
\label{data}

\begin{table}
\caption{Observational Parameters}
\begin{tabular*}{0.48\textwidth}{@{\extracolsep{\fill}}l r r r r r}
\hline
Lines & $\nu_1$ & $\nu_2$ &  $\nu_3$ & $\nu_4$ & Time \\

&(GHz)  &(GHz)  & (GHz)& (GHz)  & (min) \\
 (1) & (2) & (3) & (4) & (5)& (6)\\
 \hline
HCN, C$_2$H, CS(3-2) & 85.7 & 89.7 & 144.281 & - & 71 \\
CS(2-1), $^{13}$CO, $^{12}$CO & 93.7 & 97.7 & 109.7 & 113.7 & 61 \\
SO$_2$(?) & 101.7 & 105.7 & 172.1 & - & 40 \\
CO(2-1), CN & 227.8 & 231.8 & 243.8 & 247.8 & 40 \\
\hline
\end{tabular*}
\parbox[t]{0.48 \textwidth}{ \textit{Notes:} The tuning setups we used to obtain the spectra presented in this paper are listed here. The FTS backends installed at the IRAM-30m telescope can observe a maximum of 4$\times$4 GHz windows simultaneously. When the IRAM-30m receivers were tuned to frequencies in the 2mm band only 3 frequency windows could be used. {The tunings at 172.1, 243.8 and 247.8 GHz are not presented in this paper, as no lines were detected in those frequency ranges.} Column 1 lists some bright molecules within the target bands, and $\nu_1$-$\nu_4$ in Columns 2-5 shows the representative rest-frequency of each window used. Column 6 shows the time spent in these setups.}
\label{obssetuptable}
\end{table}

\begin{figure*}
\begin{center}
\includegraphics[height=4.5cm,angle=0,clip,trim=1.0cm 0.5cm 0.6cm 0.3cm]{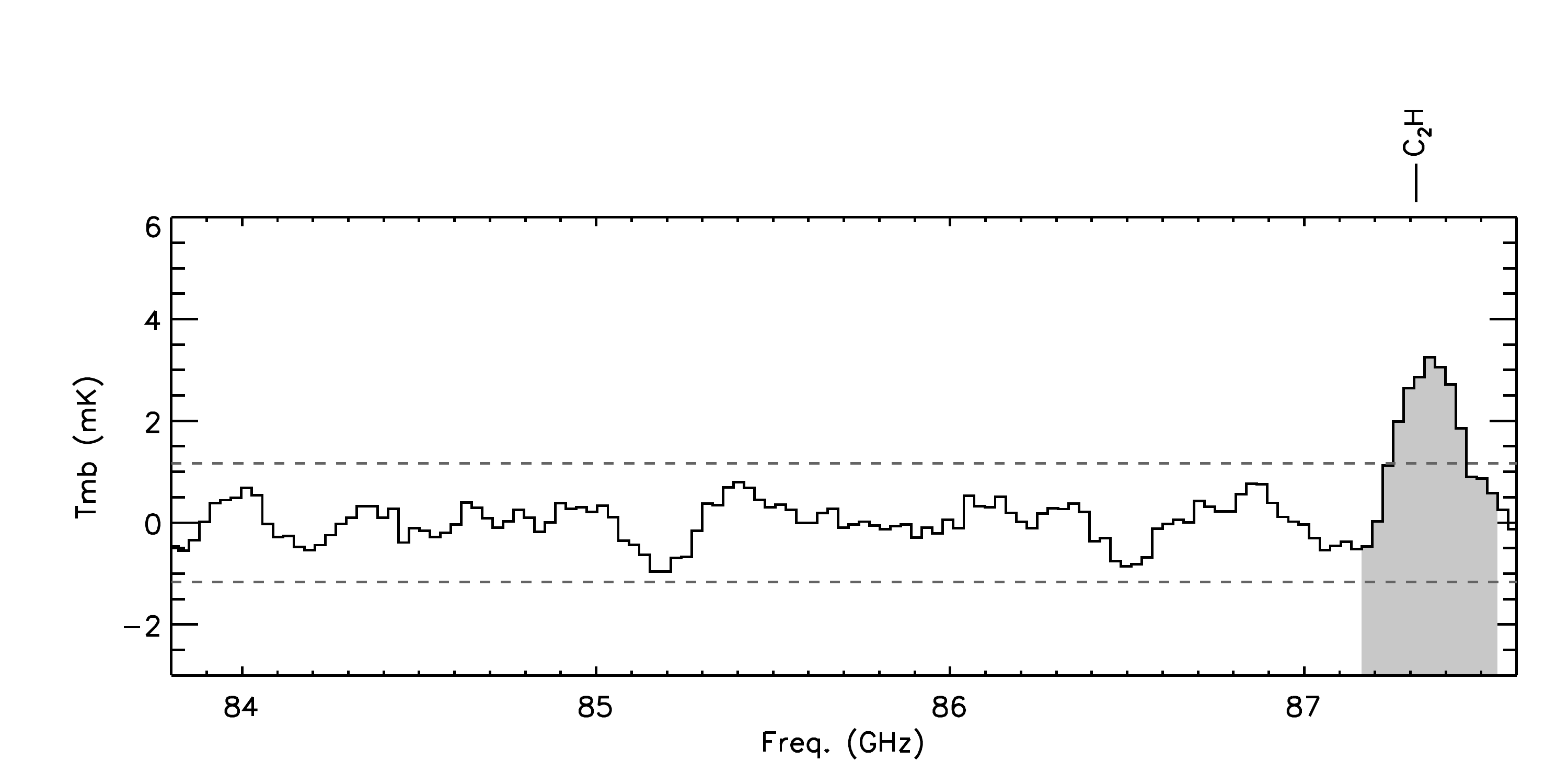}
\includegraphics[height=4.5cm,angle=0,clip,trim=3.3cm 0.5cm 0.6cm 0.3cm]{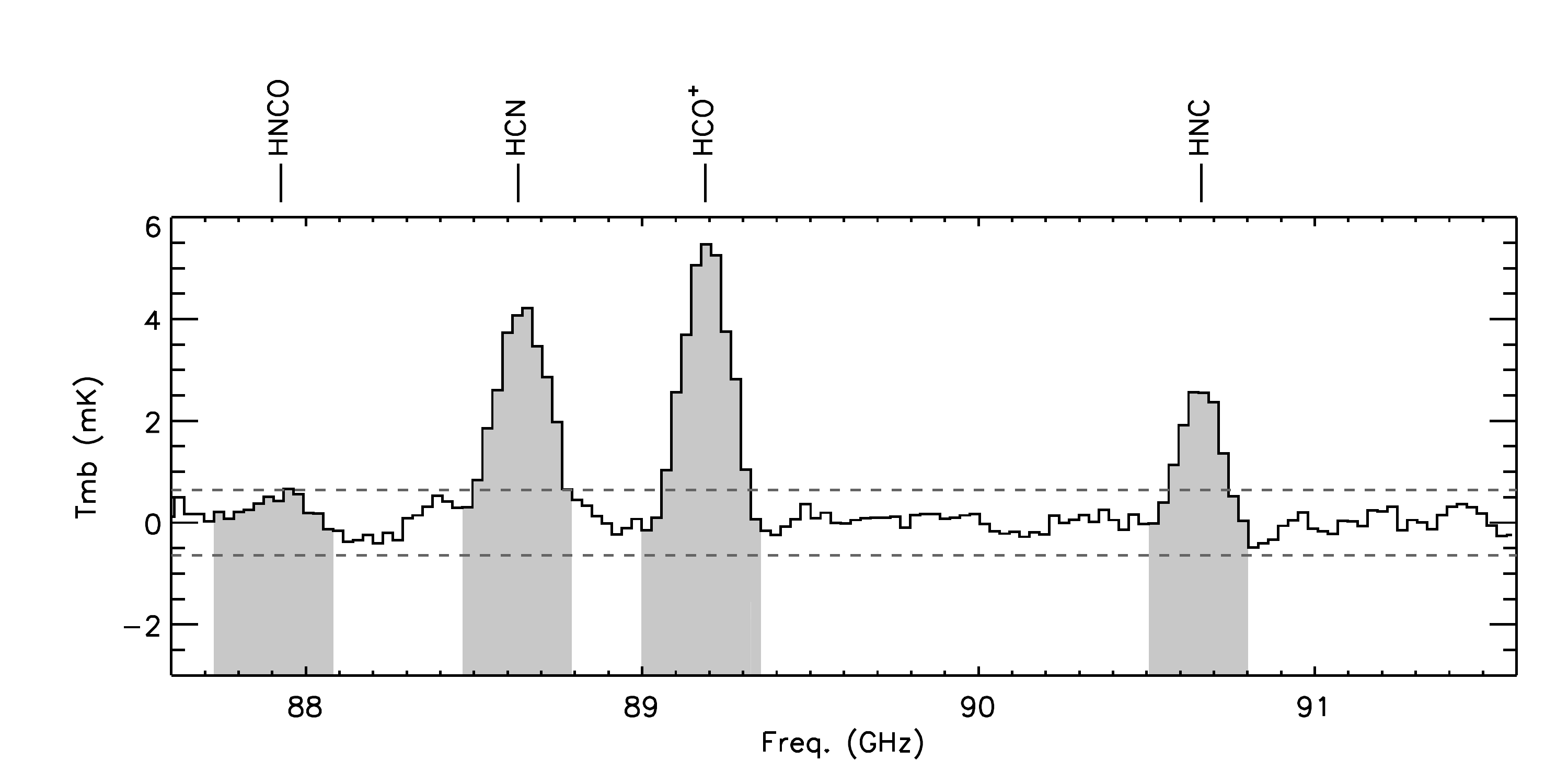}
\includegraphics[height=4.5cm,angle=0,clip,trim=1.0cm 0.5cm 0.6cm 0.3cm]{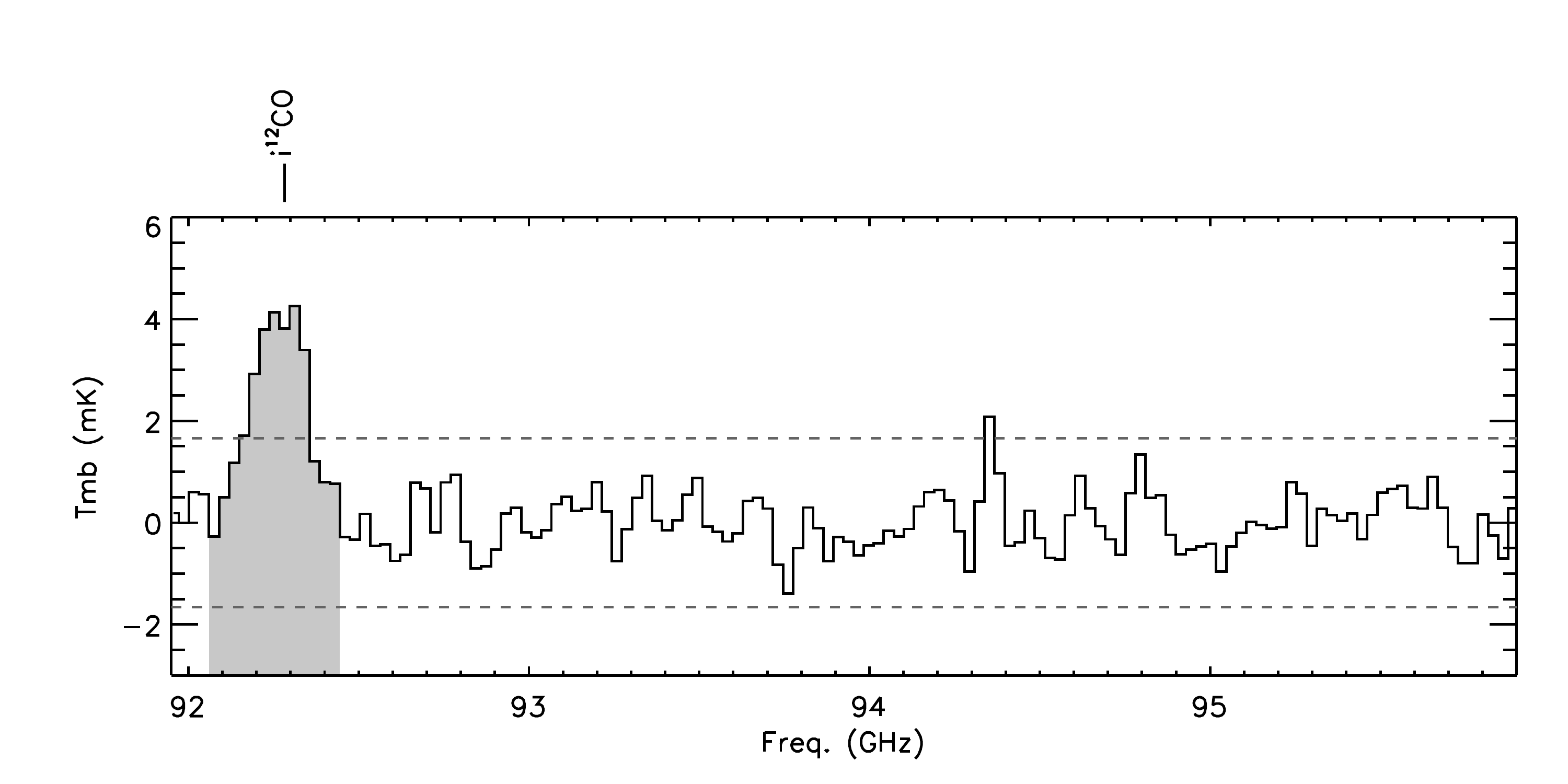}
\includegraphics[height=4.5cm,angle=0,clip,trim=3.3cm 0.5cm 0.6cm 0.3cm]{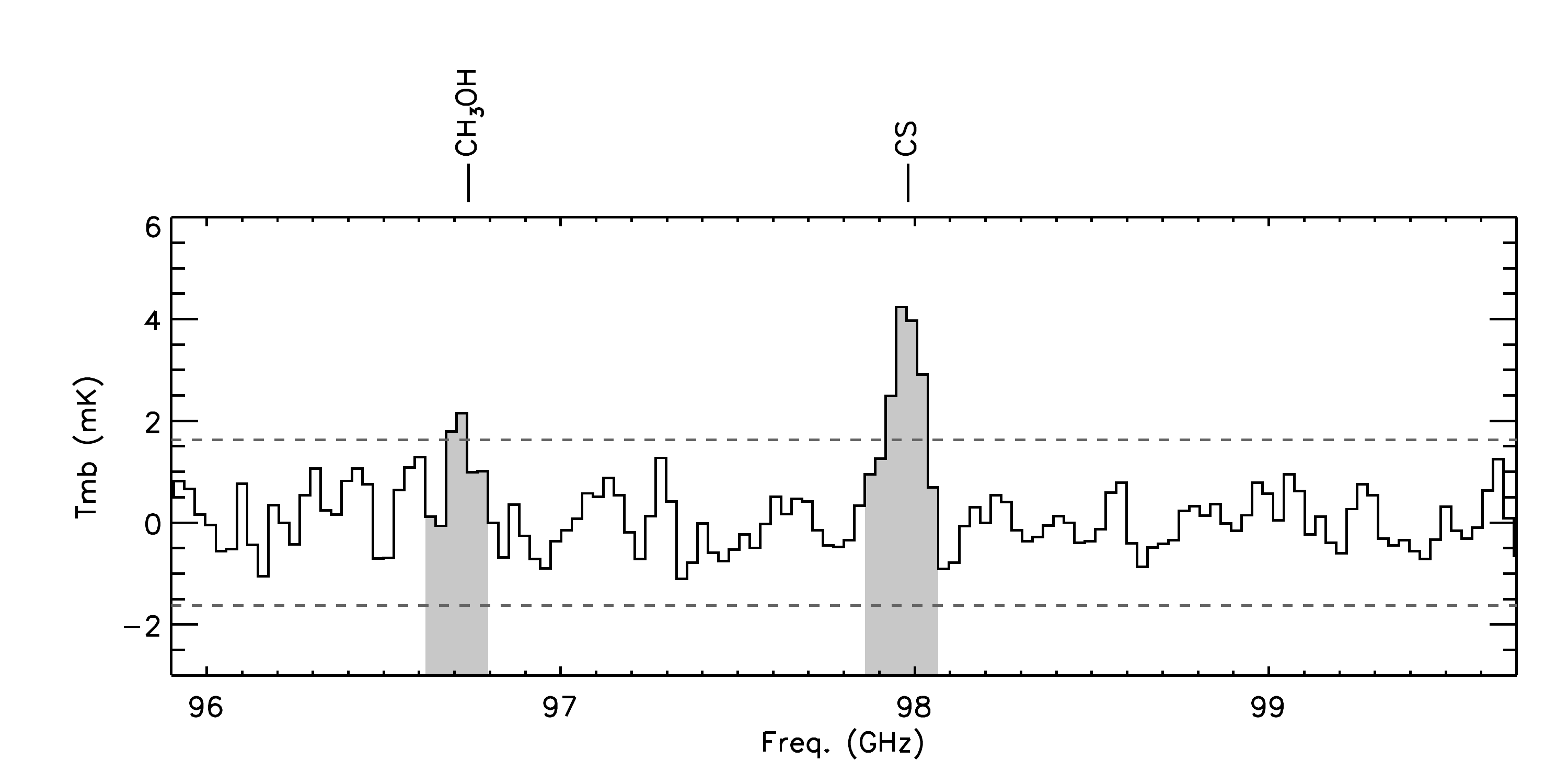}
\includegraphics[height=4.5cm,angle=0,clip,trim=1.0cm 0.5cm 0.6cm 0.3cm]{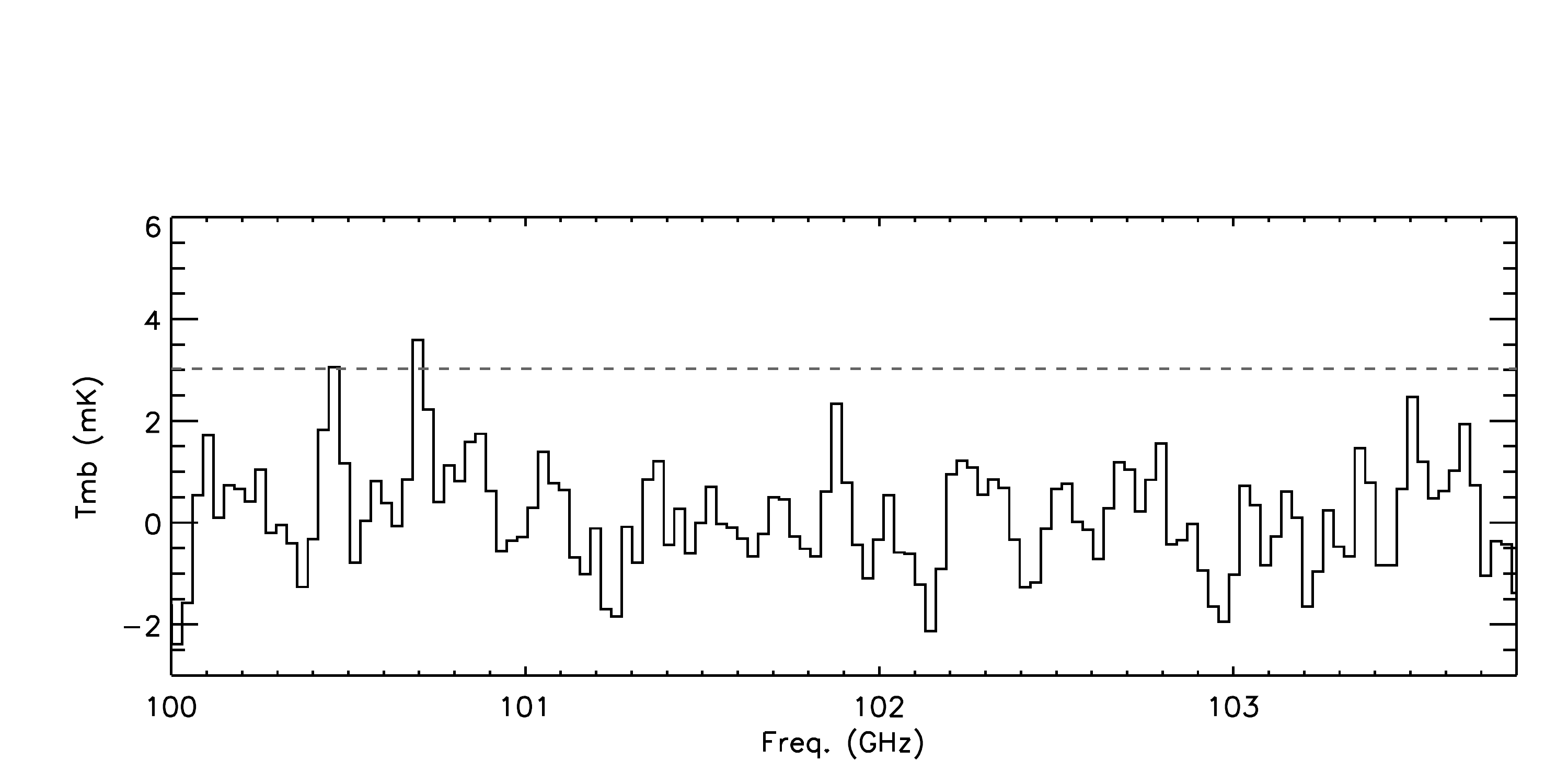}
\includegraphics[height=4.5cm,angle=0,clip,trim=3.3cm 0.5cm 0.6cm 0.3cm]{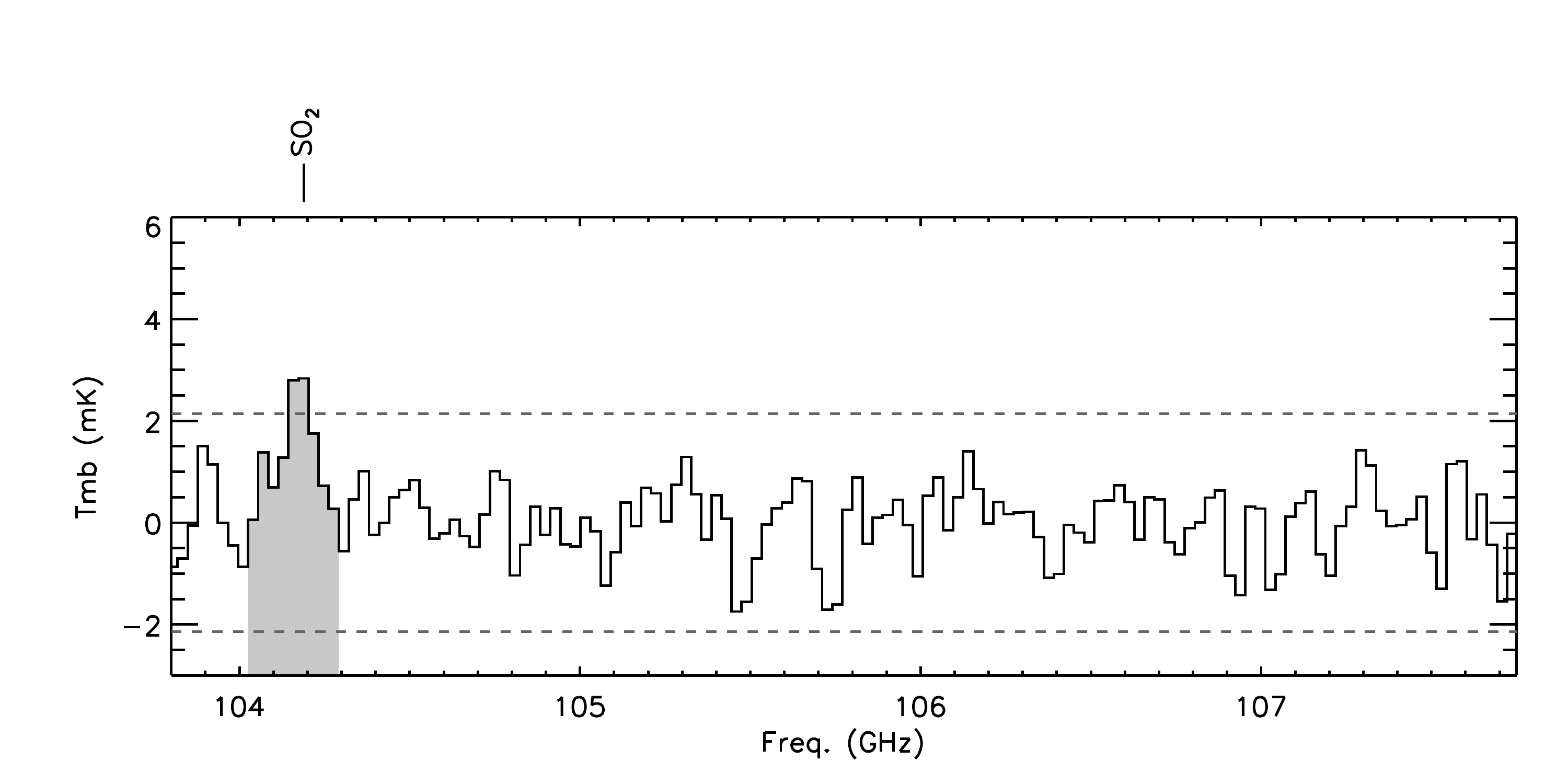}
\includegraphics[height=4.5cm,angle=0,clip,trim=1.0cm 0.5cm 0.6cm 0.3cm]{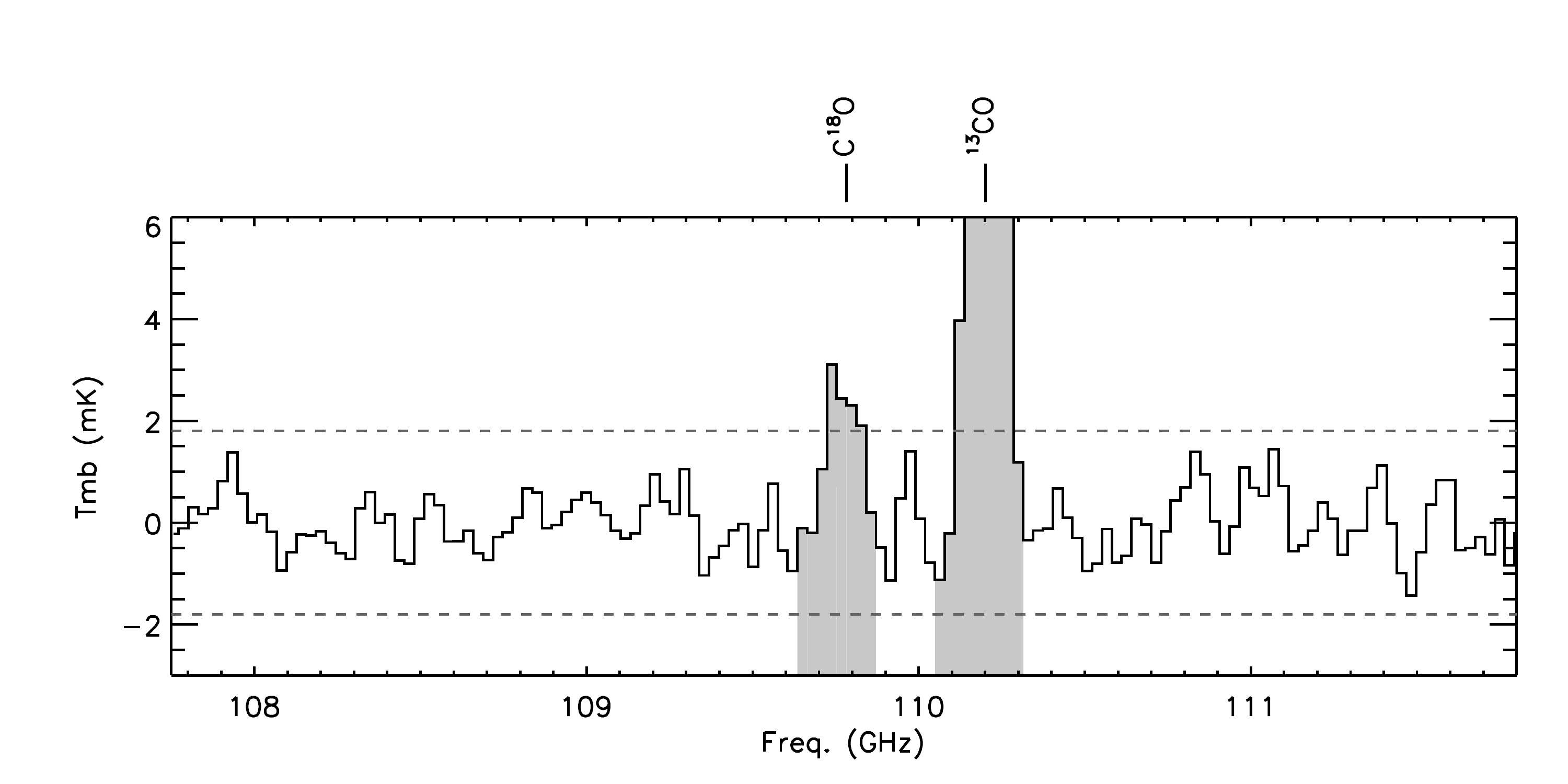}
\includegraphics[height=4.5cm,angle=0,clip,trim=3.3cm 0.5cm 0.6cm 0.3cm]{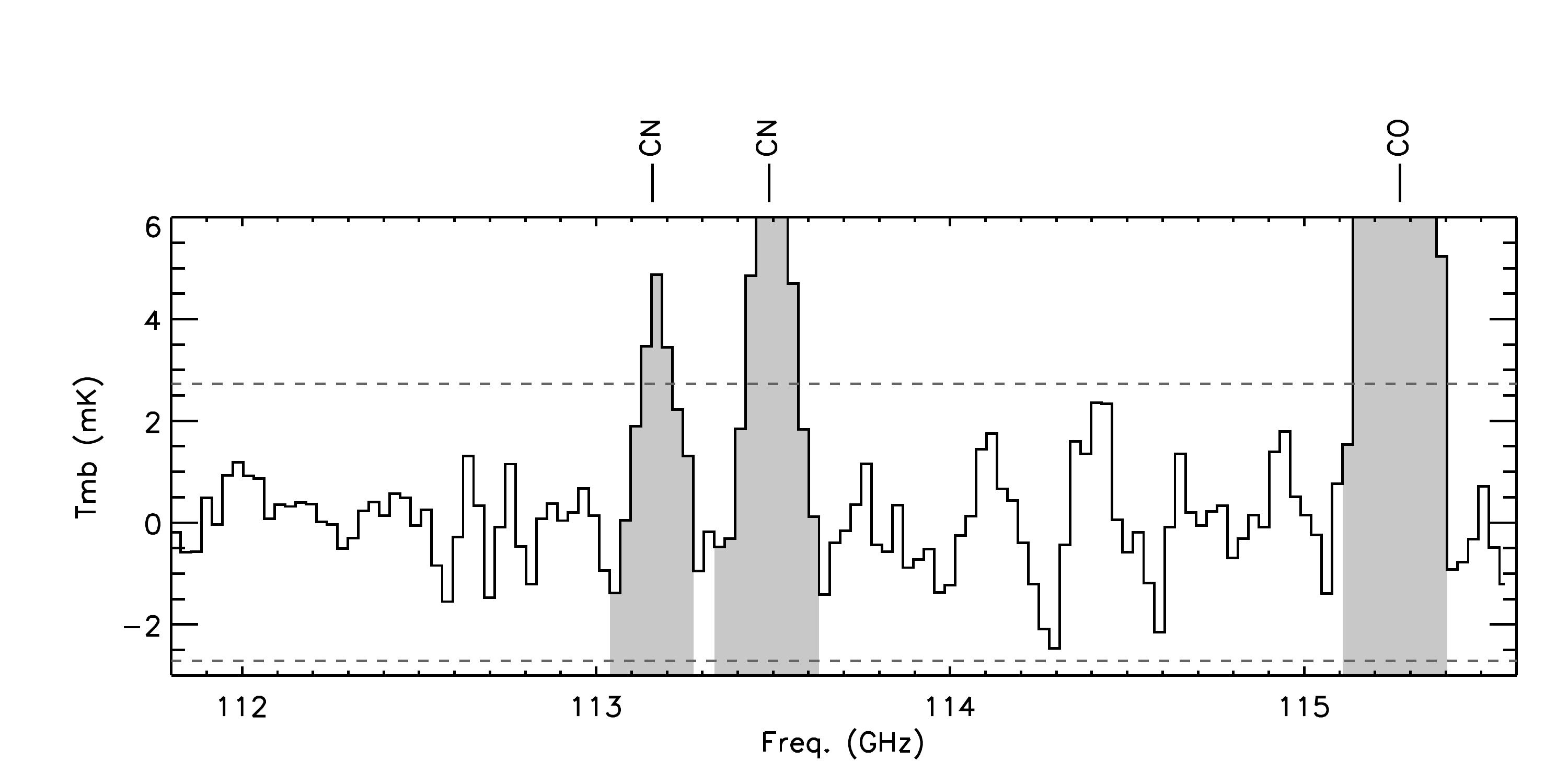}
 \end{center}
\caption{Spectral scan of the 3mm band towards Arp~157. The spectra have been binned to 100\kms channels, and grey shaded areas show the area summed over, where lines are detected. Molecules with names proceeded by an "i" are likely duplicates entering from the image band. Dashed lines show the 3$\sigma$ scatter around the fitted baseline. All plots have the same Y-axis scale (-3 to +6 mK T$_{\rm mb}$). }
 \label{linesurvey}
 \end{figure*}

\begin{figure*}
\begin{center}
\includegraphics[width=0.45\textwidth,angle=0,clip,trim=1.0cm 0.5cm 0.6cm 0.1cm]{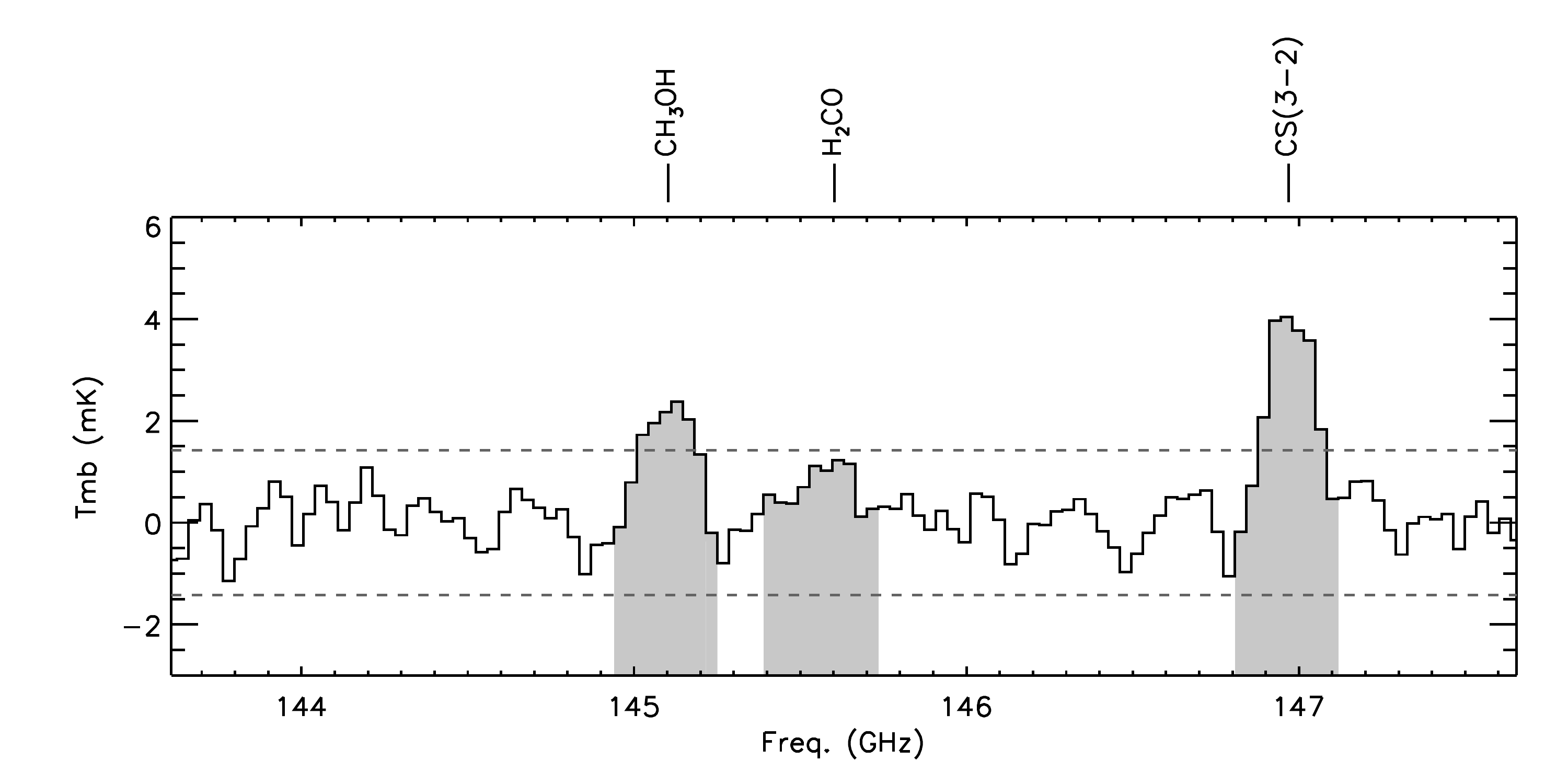}
\includegraphics[width=0.45\textwidth,angle=0,clip,trim=1.0cm 0.5cm 0.6cm 0.1cm]{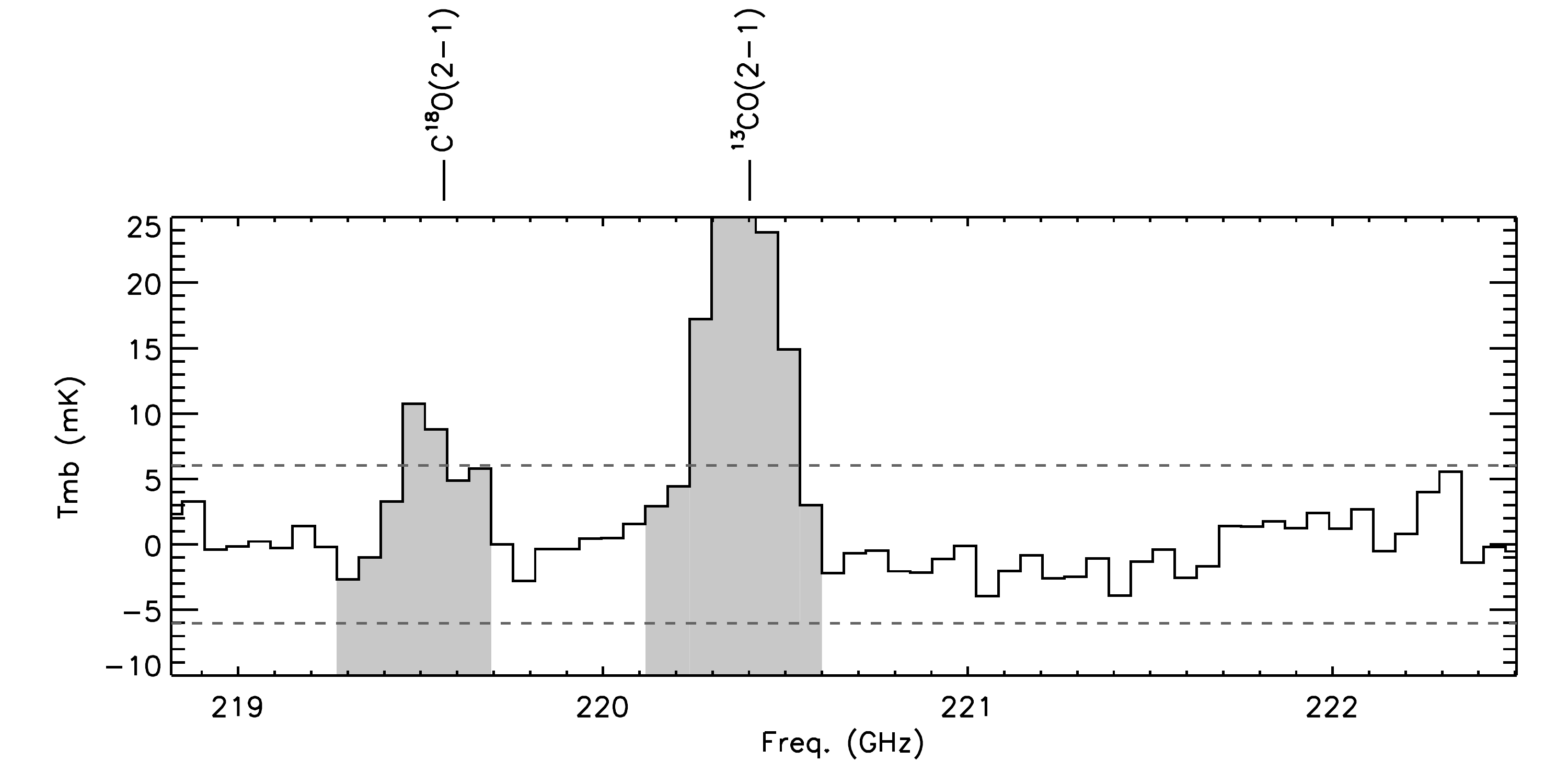}
\includegraphics[width=0.45\textwidth,angle=0,clip,trim=1.0cm 0.5cm 0.6cm 0.1cm]{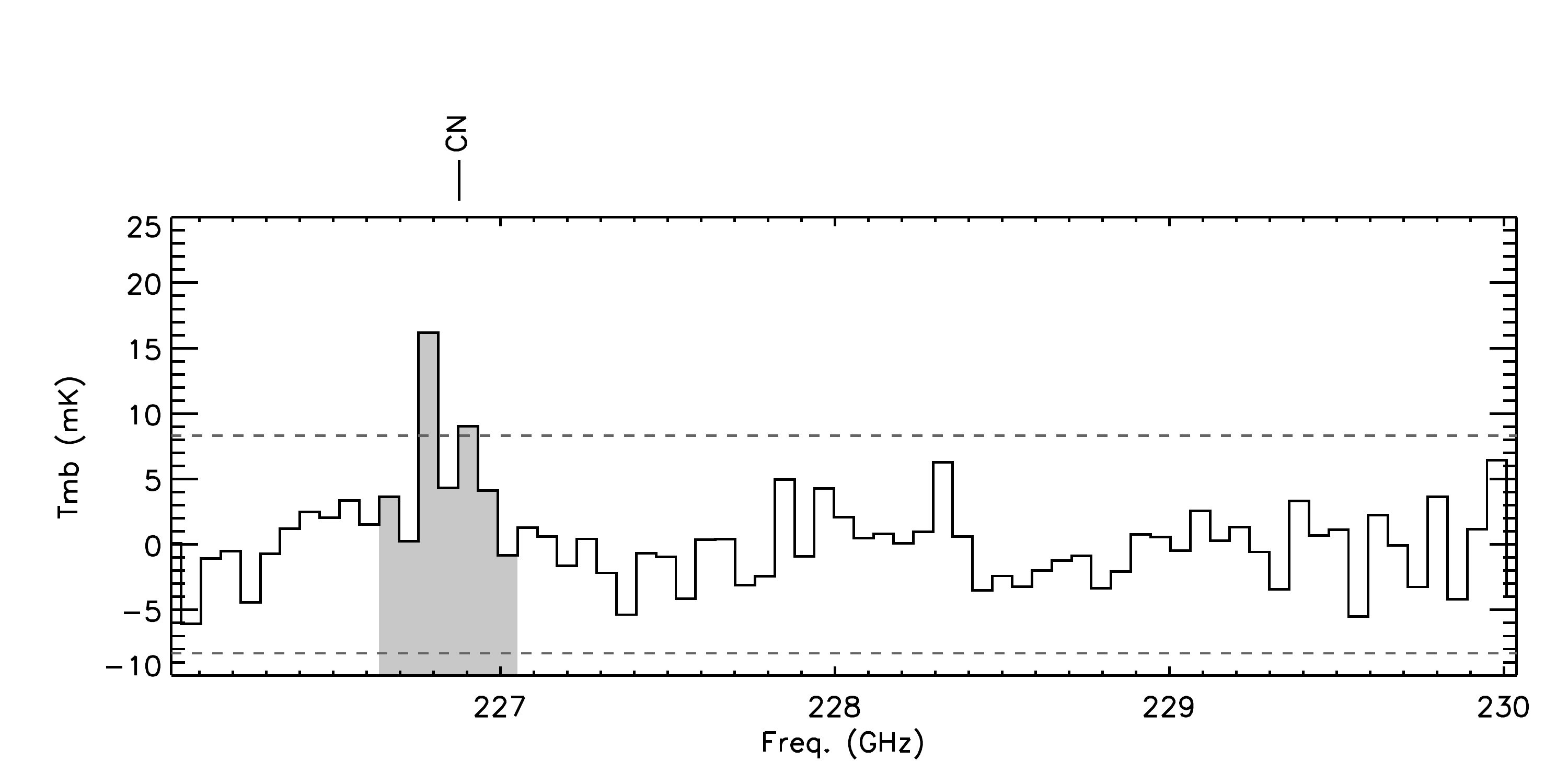}
\includegraphics[width=0.45\textwidth,angle=0,clip,trim=1.0cm 0.5cm 0.6cm 0.1cm]{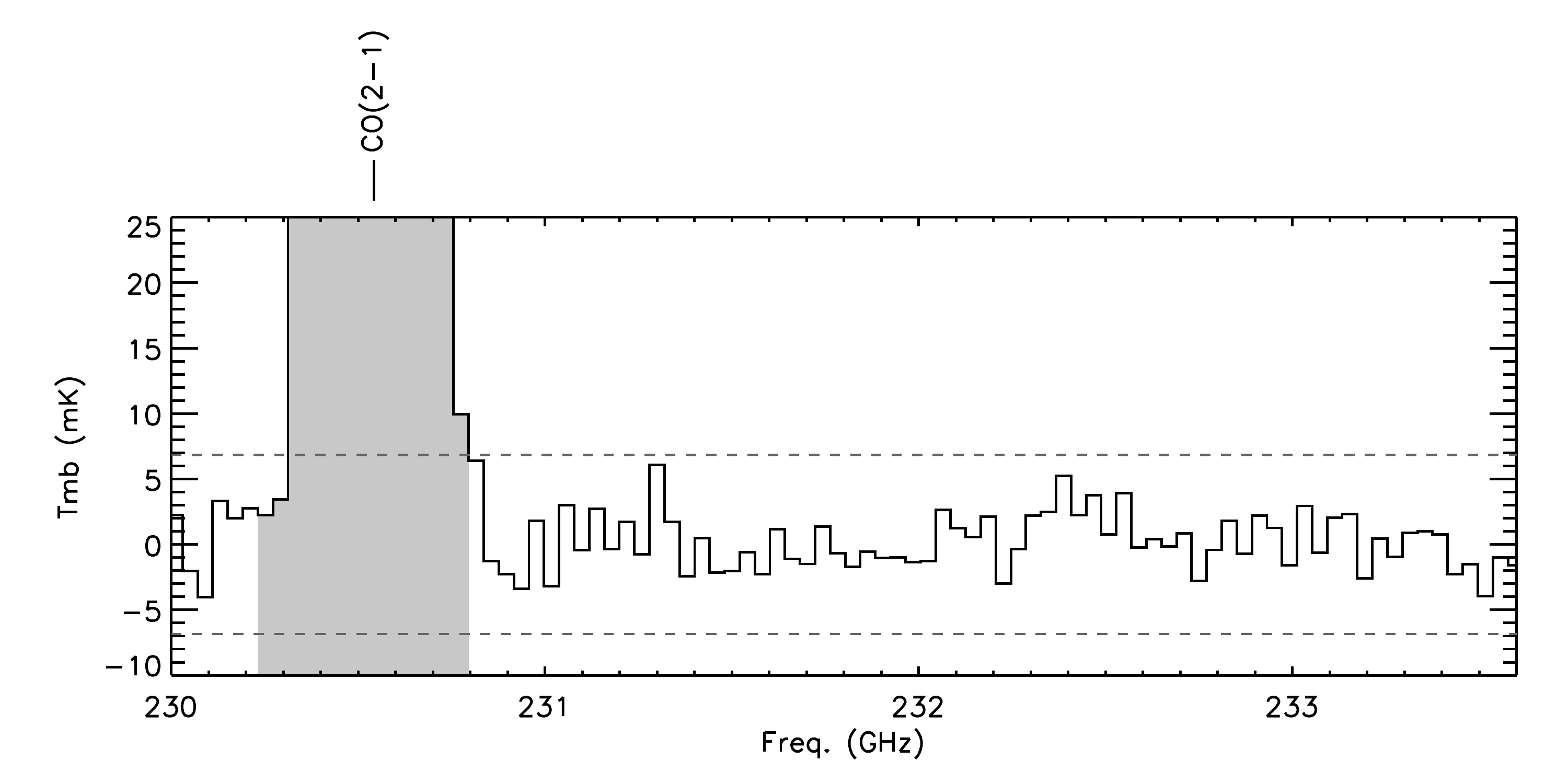}
 \end{center}
\caption{2mm and 1mm specta of Arp~157, binned to 100\kms channels. Details as in Figure \protect \ref{linesurvey}. }
 \label{linesurvey_highfreq}
 \end{figure*}
 
The IRAM 30-m telescope at Pico Veleta, Spain, was used during January 2012 to observe Arp~157 in the 3 and 2mm bands (and in October 2012 to observe the 1mm band). The beam full width at half-maximum (FWHM) of the IRAM-30m ranges from 27\arcsec\ at 85 GHz to 11\arcsec\ at 230 GHz.  The EMIR receiver was used for observations in the wobbler switching mode, with reference positions offset by $\pm$100\arcsec\ in azimuth. The FTS back-end processor gave a total bandwidth of $\approx$16 GHz in double-sideband mode, and a raw spectral resolution of 200 kHz ($\approx$0.6 \kms\ at $\lambda$=3mm, $\approx$0.4 \kms\  at $\lambda$=2mm, $\approx$0.3 \kms\  at $\lambda$=1mm). 

The system temperatures ranged between {80 and 102}~K at 3~mm, {87 and 103}~K at 2~mm and {230 and 242}~K at 1~mm.
 The pointing was checked every 2 hours on a nearby planet or bright quasar, and the focus was checked at the beginning of each night as well as after sunrise or more often if a suitable planet was available. 
 Table \ref{obssetuptable} shows the frequency setups used (and the time spent observing each).
 
The individual $\approx$6-min scans were inspected, and baselined, either with a simple constant (zero-order) baseline, a linear (first-order) baseline, or a second order polynomial, depending on the scan. The FTS backend at the IRAM 30m suffers from platforming problems, where well defined channel ranges have large flux offsets from one another. These were dealt with by individually fitting baselines to the affected regions. Scans with poor quality baselines or other problems were discarded. The good scans were averaged together, weighted by the inverse square of the system temperature. 

We convert the spectra from the observed antenna temperature (T$_a^*$) to main beam temperature (T$_{\rm mb}$) by dividing by the ratio of the beam and forward efficiencies ($\eta=$  B$_{\rm eff}$/F$_{\rm eff}$) in the standard manner, as in \cite{2013MNRAS.433.1659D}. The resultant spectra are shown in Figure \ref{linesurvey}.

Where lines were detected, integrated intensities were computed by direct summation of the spectrum over velocity.
The line widths summed over (presented in Table \ref{obstable}) were estimated by fitting gaussians to the lines. If we had instead used these gaussian fits to estimate the integrated intensities, this would not have changed our conclusions. 

Table \ref{obstable} lists the observed line parameters. Uncertainties were calculated for each 4GHz band separately, as in \citealt{2012MNRAS.421.1298C}. Bands observed with the same setup at the same time can thus have different RMS noise estimates, especially if the band is near the edge of the possible tuning range, or if noise spikes or weak emission from formally undetected species are present. When the integrated intensity of a suspected line is greater than three times its own uncertainty (including the uncertainty from estimating the baseline level), we include the line as a detection. 

The source-averaged brightness temperature (T$_{\rm B}$) can be estimated from the measured main beam brightness temperature (T$_{\rm mb}$). In the approximation of a Gaussian source distribution of size major axis size $\theta_{\rm s,maj}$ and minor axis size $\theta_{\rm s,min}$, observed with a Gaussian beam of size $\theta_{b}$ the below equations correct for the dilution effect due to the coupling between the source and the telescope beam \citep[e.g.][]{2006ApJS..164..450M}.

\begin{equation}
T_{\rm B} = \frac{T_{\rm mb}}{\phi_{\rm ff}},
\label{beamdilu}
\end{equation}
where $\phi_{\rm ff}$ is the beam filling factor defined as
\begin{equation}
\phi_{\rm ff} = \frac{\theta_{\rm s,maj}\theta_{\rm s,min}}{\theta_{\rm s,maj}\theta_{\rm s,min}+\theta_{b}^2}
\end{equation}

Here the source size is estimated using $\theta_{\rm s,maj}$=6\arcsec\ and $\theta_{\rm s,min}$=2\arcsec, as revealed from the interferometric maps of \cite{2003MNRAS.346..424B}. The source-averaged brightness temperatures are listed in Table \ref{obstable}. The real extent of each emitting component is likely to be different, and this is one of the main uncertainties in our resulting quantities.

\begin{table*}
\caption{Molecular transitions detected.}
\begin{tabular*}{0.8\textwidth}{@{\extracolsep{\fill}}l r r r r r r r r r}
\hline
Transition & Freq & Beam & Peak & RMS & $\Delta V$ & $\int T_{\rm mb}\, \delta V$ & Error & $\int T_b\, \delta V$ & Error \\
  &(GHz)  & (\arcsec)& (mK)  & (mK)  & (\kms) & (K \kms) & (K \kms) & K \kms & K \kms  \\
 (1) & (2) & (3) & (4) & (5) & (6) & (7) & (8) & (9) & (10)\\
 \hline
 C$_2$H & 87.316 & 28.2 & 3.25 & 0.39 & 549 & 2.04 & 0.16 & 137.03 & 10.65\\
HNCO & 87.925 & 28.0 & 0.66 & 0.21 & 500 & 0.30 & 0.08 & 19.71 & 5.45\\
HCN & 88.631 & 27.7 & 4.21 & 0.21 & 500 & 2.79 & 0.09 & 181.78 & 5.54\\
HCO$^+$ & 89.188 & 27.6 & 5.46 & 0.21 & 500 & 3.01 & 0.08 & 193.28 & 5.30\\
HNC & 90.663 & 27.1 & 2.56 & 0.21 & 438 & 1.18 & 0.08 & 73.31 & 4.77\\
CH$_3$OH & 96.739 & 25.4 & 2.16 & 0.54 & 237 & 0.55 & 0.13 & 29.77 & 7.12\\
CS & 97.981 & 25.0 & 4.24 & 0.54 & 310 & 1.38 & 0.16 & 73.20 & 8.30\\
SO$_2$? & 104.190 & 23.5 & 2.83 & 0.71 & 310 & 0.91 & 0.21 & 42.67 & 9.76\\
C$^{18}$O & 109.782 & 22.3 & 3.11 & 0.60 & 280 & 0.65 & 0.16 & 27.68 & 6.77\\
$^{13}$CO & 110.201 & 22.2 & 15.24 & 0.60 & 313 & 4.90 & 0.17 & 205.69 & 7.02\\
CN$_{1-4}$ & 113.157 & 21.6 & 4.87 & 0.91 & 272 & 1.19 & 0.23 & 47.25 & 9.19\\
CN$_{5-9}$ & 113.499 & 21.5 & 7.79 & 0.91 & 335 & 2.55 & 0.26 & 101.09 & 10.41\\
CO & 115.271 & 21.2 & 280.91 & 0.91 & 335 & 102.59 & 0.26 & 3937.99 & 10.09\\
CH$_3$OH & 144.195 & 16.8 & 2.38 & 0.47 & 299 & 0.73 & 0.12 & 17.75 & 2.94\\
H$_2$CO & 145.603 & 16.6 & 1.23 & 0.47 & 299 & 0.55 & 0.12 & 13.15 & 2.92\\
CS & 146.969 & 16.5 & 4.04 & 0.47 & 299 & 1.46 & 0.12 & 34.46 & 2.76\\
C$^{18}$O & 219.564 & 10.8 & 10.77 & 2.01 & 249 & 2.22 & 0.49 & 23.96 & 5.33\\
$^{13}$CO & 220.403 & 10.8 & 30.24 & 2.01 & 307 & 10.25 & 0.55 & 109.65 & 5.85\\
CN$_{5-9}$ & 226.876 & 10.5 & 16.18 & 2.77 & 250 & 3.06 & 0.67 & 31.03 & 6.79\\
CO & 230.542 & 10.3 & 564.34 & 2.28 & 313 & 187.55 & 0.50 & 1842.07 & 4.96\\
\hline
\end{tabular*}
\parbox[t]{0.8 \textwidth}{ \textit{Notes:}  Column 1 lists the detected transition, and its frequency is shown in Column 2. Column 3 shows the beam size of the IRAM-30m telescope at this frequency, estimated by linearly interpolating the measured beam sizes tabulated at the IRAM-30m website$^{*}$. Columns 4 and 5 show the peak main beam brightness temperature of the line, and the RMS around the baseline calculated in line free regions. Column 6 shows the line width. The observed integrated main beam intensity is shown in Column 7, and its error is shown in Column 8. Column 9 and 10 show this integrated intensity in beam temperature units (i.e. corrected for beam dilution) using Equation \ref{beamdilu}, and a source size of 6$\times$2 arcseconds, estimated from the interferometric imaging of \cite{2003MNRAS.346..424B}.}
\label{obstable}
\end{table*}
 
 \section{Results}
 \label{results}
  
\footnotetext{http://www.iram.es/IRAMES/mainWiki/EmirforAstronomers - accessed 23/04/2013}
 \subsection{Line detections} 
\label{linesfound}
In this relatively shallow line survey we detect 19 different lines, from 12 different molecular species. All of these molecules are well detected, and line identification is relatively simple as the species are common in the ISM of other nearby galaxies. 

The exception is the line provisionally identified as SO$_2$ in Figure \ref{linesurvey}. The detection of this line is reasonably significant ($\approx$5$\sigma$), and no potentially contaminating bright lines exist in the image band at this location. 
SO$_2$ is the strongest line present in this spectral region; however it is relatively rarely detected in distant galaxies, and the centre of the detected line is {$\approx$500 \kms\ away from the expected rest-frequency}. 
In addition, an SO$_2$ feature of similar expected intensity should be present at 146.6 GHz, which is not observed in Figure \ref{linesurvey_highfreq}.
Signal from SO$_2$ could be blended with emission from S$^{17}$O; however as SO lines themselves are not detected, any emission from isotopologues should be weak.
The abundance of SO$_2$ has been shown to increase with metallicity \citep{2012MNRAS.424.2646B}, so its high abundance here may be related to the relatively high metallicity of the gas in this system (around solar; \citealt{1991ApJ...381..409S}). If this were the case, an increase in the fractional abundance of other molecules (such as HCN) would also be expected, and this is not observed (see Section \ref{chem_abund_gals}). 
{Because of the uncertainty about the true identification of this line we do not use it in the rest of the analysis in this paper.}

\subsection{Literature comparison}

Arp~157 has been observed in the $^{12}$CO and HCN lines by other authors, and in this section we compare the measured fluxes.
{All these fluxes have been converted to Janskies using the applicable point source conversion from the main beam temperature to flux density, which for our data is taken from the tables on the IRAM website\footnote{http://www.iram.es/IRAMES/mainWiki/Iram30mEfficiencies - accessed 23/04/2013}.)}

\cite{1992ApJ...387L..55S} observed this object with the IRAM-30m, and reported a detection of $^{12}$CO(1-0) and HCN(1-0) with integrated fluxes of 211.2 and 14.9 Jy \kms, respectively (no error bars given). 
Here we detect both transitions well, and our derived fluxes are 473$\pm$1 and 13.8$\pm$0.4 Jy \kms\ respectively. The HCN detection of \cite{1992ApJ...387L..55S} is consistent with ours within 3$\sigma$, however the $^{12}$CO measurement is not. More recent measurements by \cite{2001ApJ...550..104Y} with OVRO are more consistent however, reporting a $^{12}$CO flux of 404$\pm$61 Jy \kms\ for this object. Overall we thus consider that our measurements agree with those made by other authors. 

\subsection{Line profiles}

\begin{figure}
\begin{center}
\includegraphics[width=0.5\textwidth,angle=0,clip,trim=0cm 0cm 0cm 0cm]{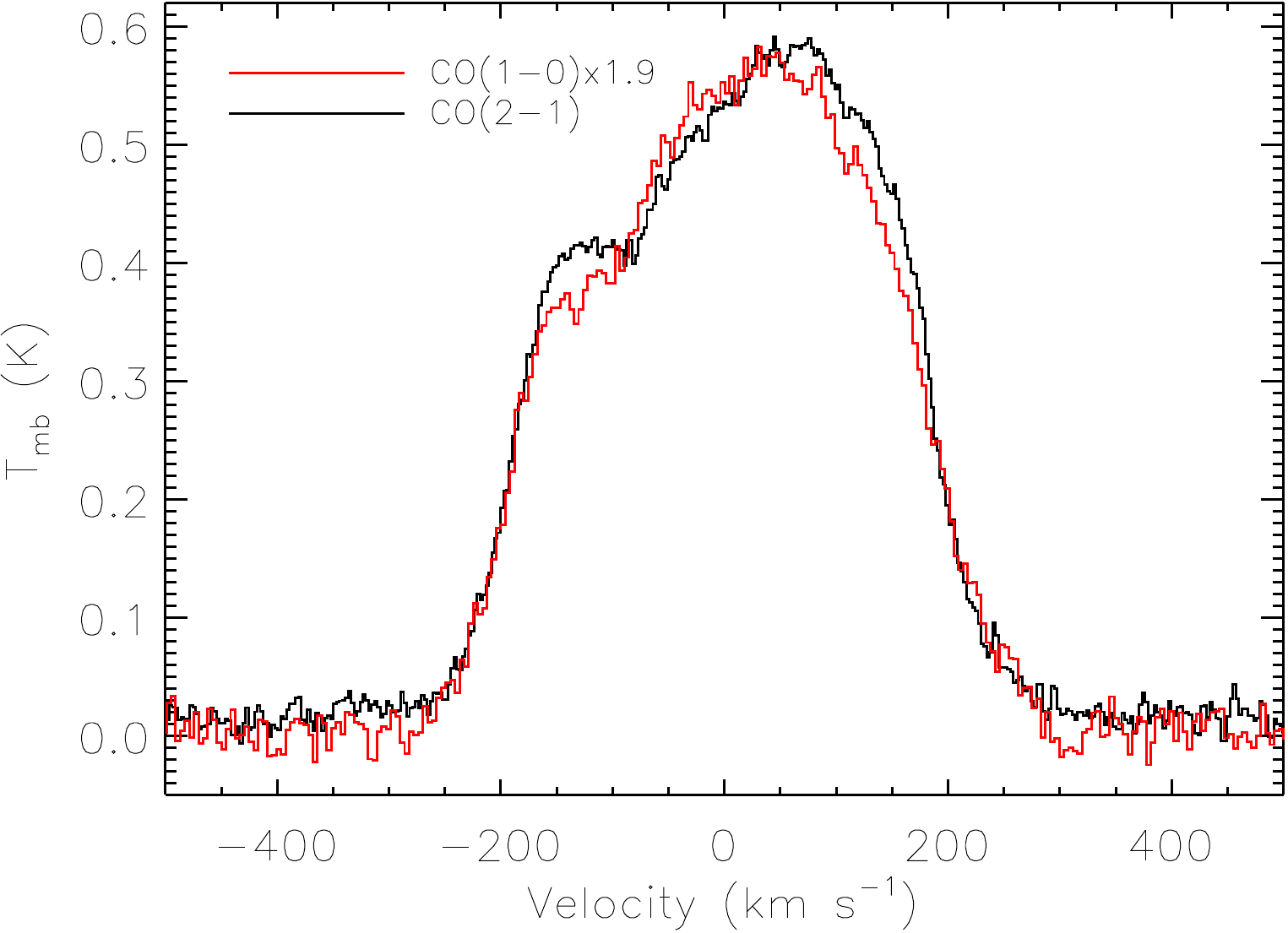}
 \end{center}
\caption{\textit{Top:} $^{12}$CO line profiles in Arp~157, at 4.2 and 2.6 \kms\ resolution for $^{12}$CO(1-0) (red line) and $^{12}$CO(2-1) (black line), respectively. The velocity axis is relative to the galaxy systemic, here taken as 2258 \kms.The $^{12}$CO(1-0) spectrum has been multiplied by 1.9 to better show the slight difference in profile shape between the lines.  }
\label{cospec}
 \end{figure}

As previous authors have shown, molecular gas line profiles in this source are not gaussian. In this work for the brightest lines we have the signal-to-noise to quantify this. Figure \ref{cospec} shows the $^{12}$CO(1-0) and $^{12}$CO(2-1) line profiles at $<$5\kms\ resolution. The profile is clearly not gaussian, with more luminous emission at redshifted velocities.   Even in these two closely related lines the profiles differ, with the $^{12}$CO(1-0) emission being more centrally peaked, and the blue wing of the $^{12}$CO(1-0) being less luminous. Other gas tracers show even bigger differences in their line profile shapes, with some (e.g HCO$^+$) being double horned and almost dropping to zero at the galaxy systemic velocity. These velocity components do not correspond to the two nuclei of the system, as molecular gas only exists around the starburst nucleus. {The total velocity width of all the lines (without hyperfine components) are consistent within our errors, and match well the velocities seen in \hi\ absorption studies of the ring around the starburst nucleus \citep{2003MNRAS.346..424B}}. 

\begin{figure*}
\begin{center}
\includegraphics[width=0.45\textwidth,angle=0,clip,trim=0cm 2.1cm 0cm 1.0cm]{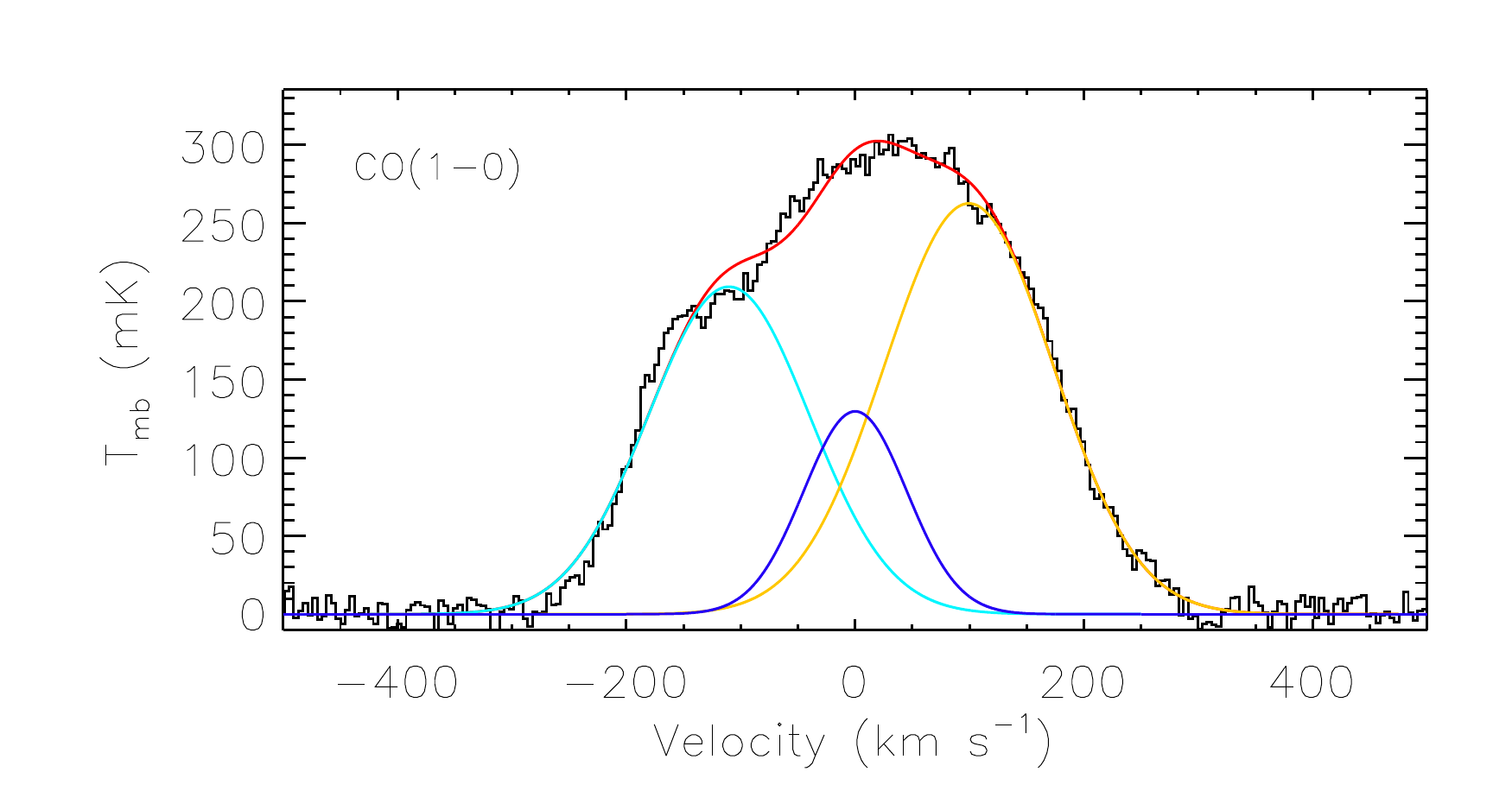}
\includegraphics[width=0.45\textwidth,angle=0,clip,trim=0cm 2.1cm 0cm 1.0cm]{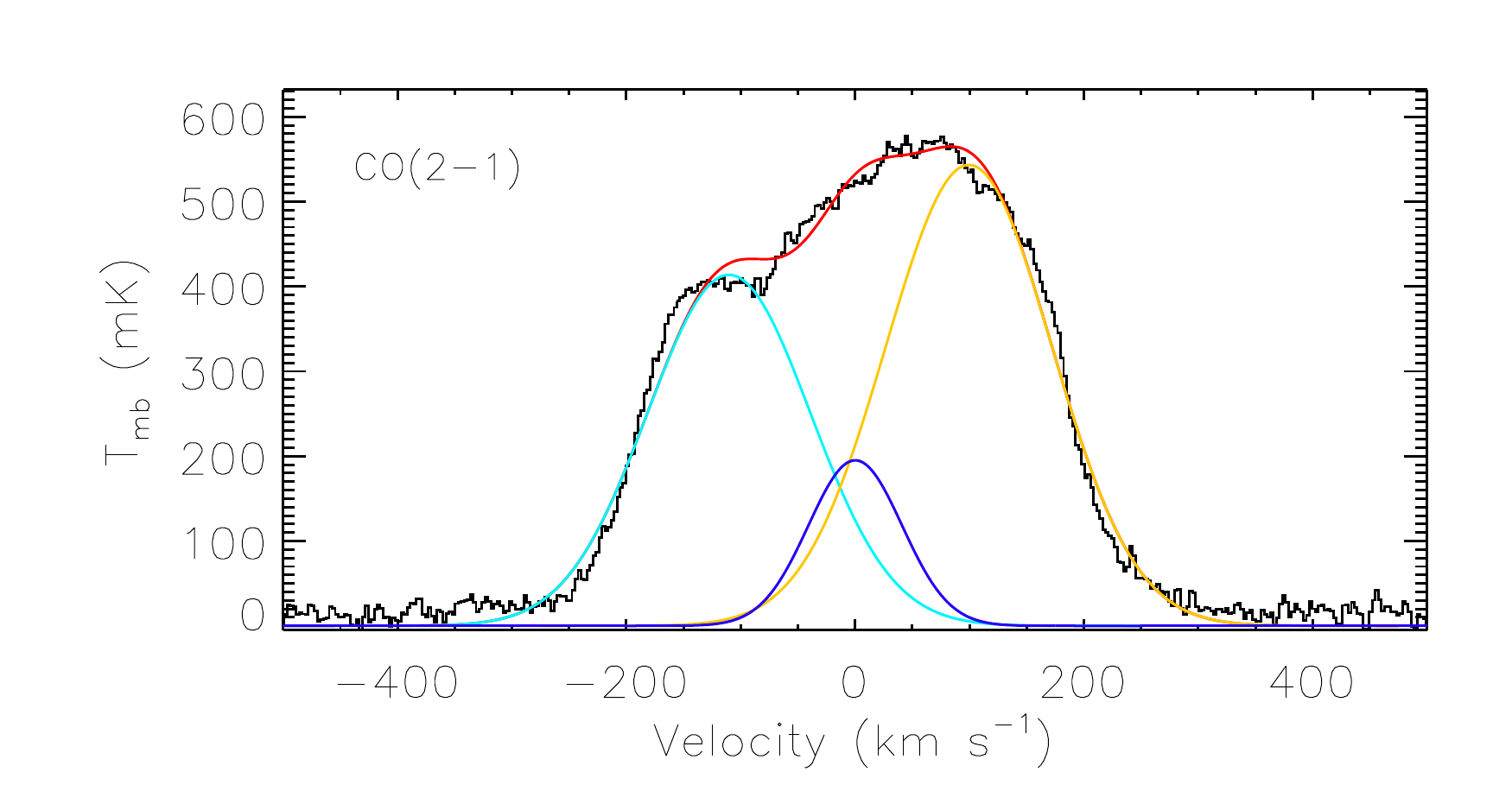}\\
\includegraphics[width=0.45\textwidth,angle=0,clip,trim=0cm 2.1cm 0cm 1.0cm]{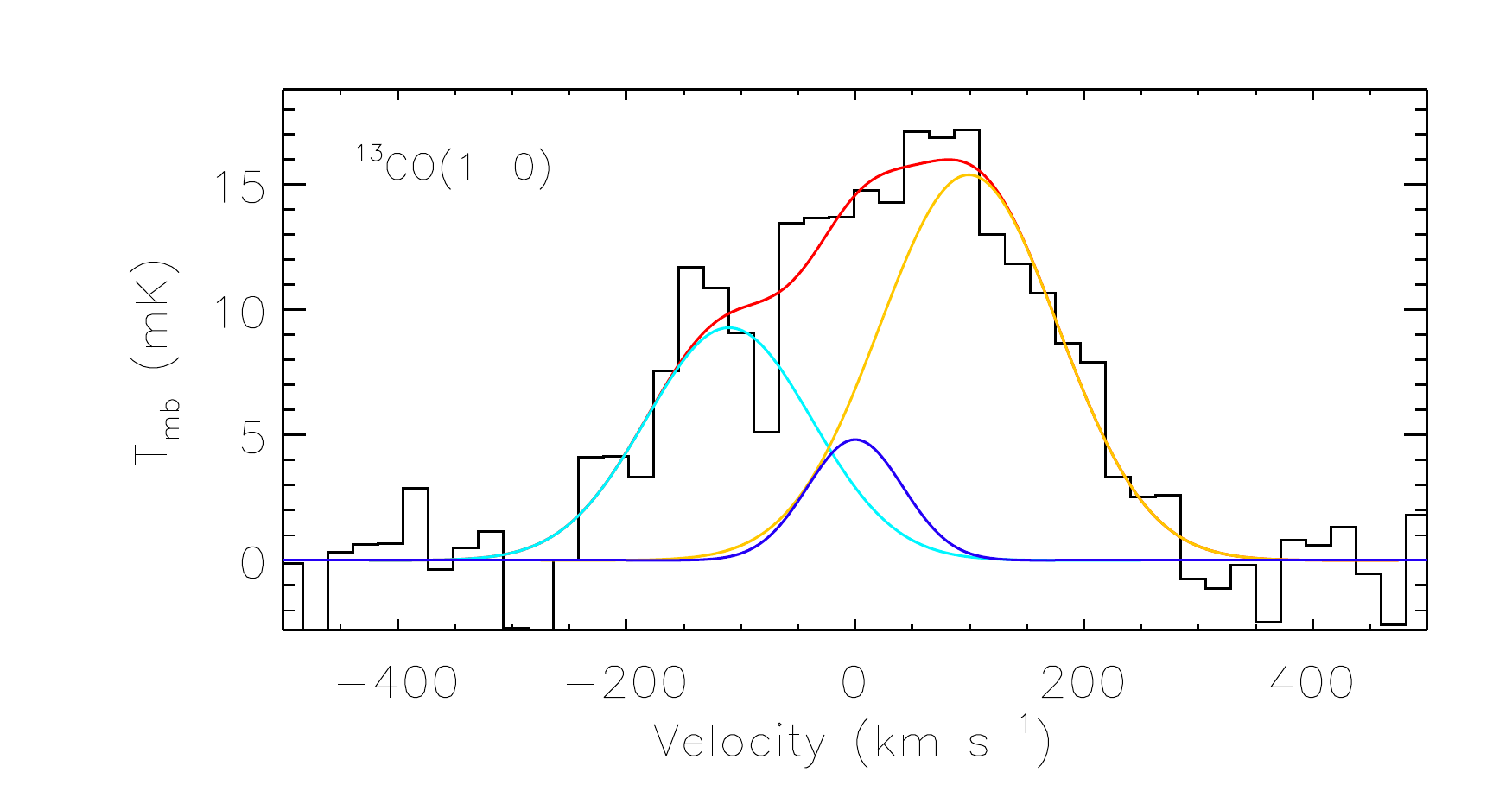}
\includegraphics[width=0.45\textwidth,angle=0,clip,trim=0cm 2.1cm 0cm 1.0cm]{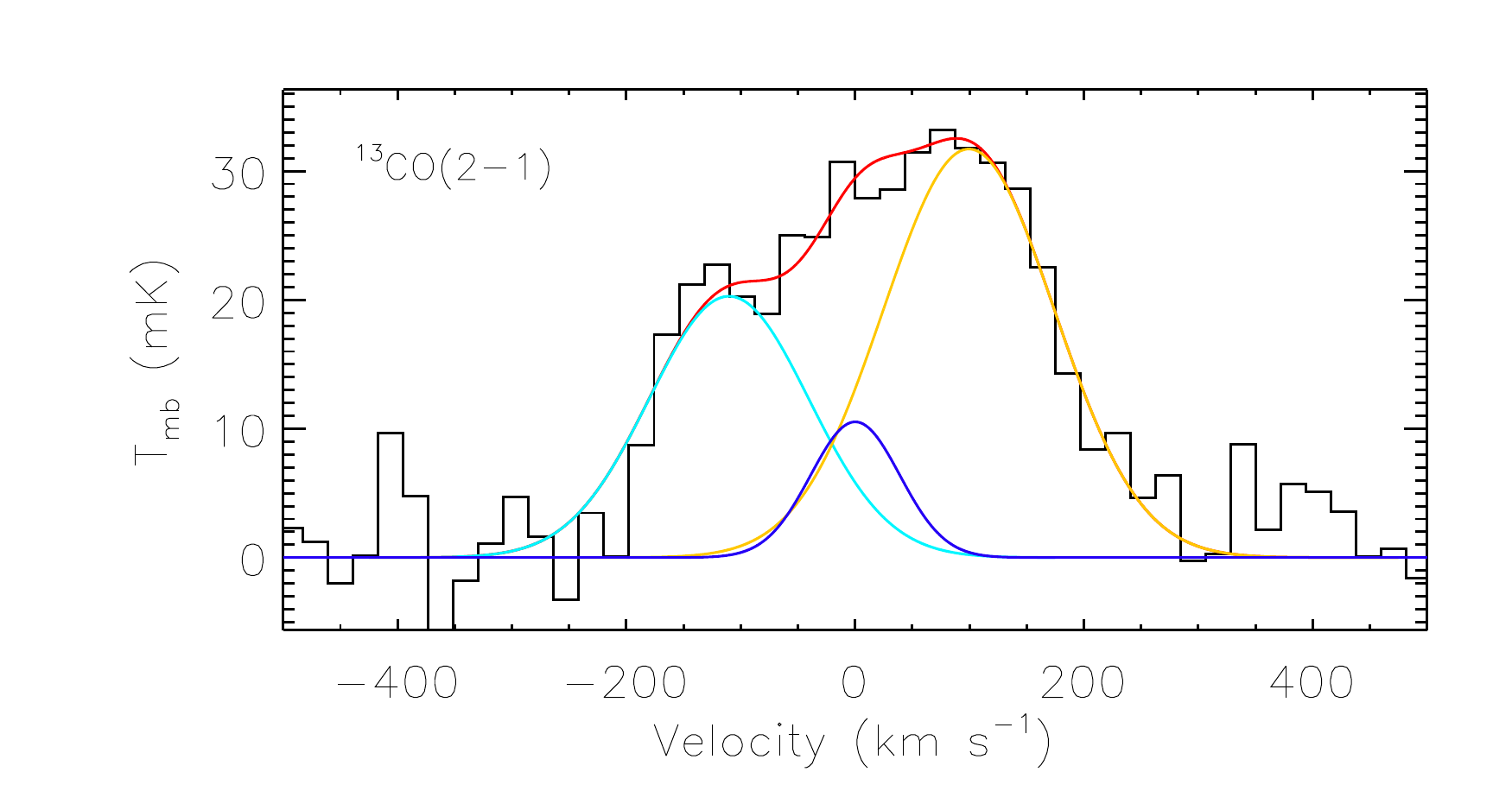}\\
\includegraphics[width=0.45\textwidth,angle=0,clip,trim=0cm 2.1cm 0cm 1.0cm]{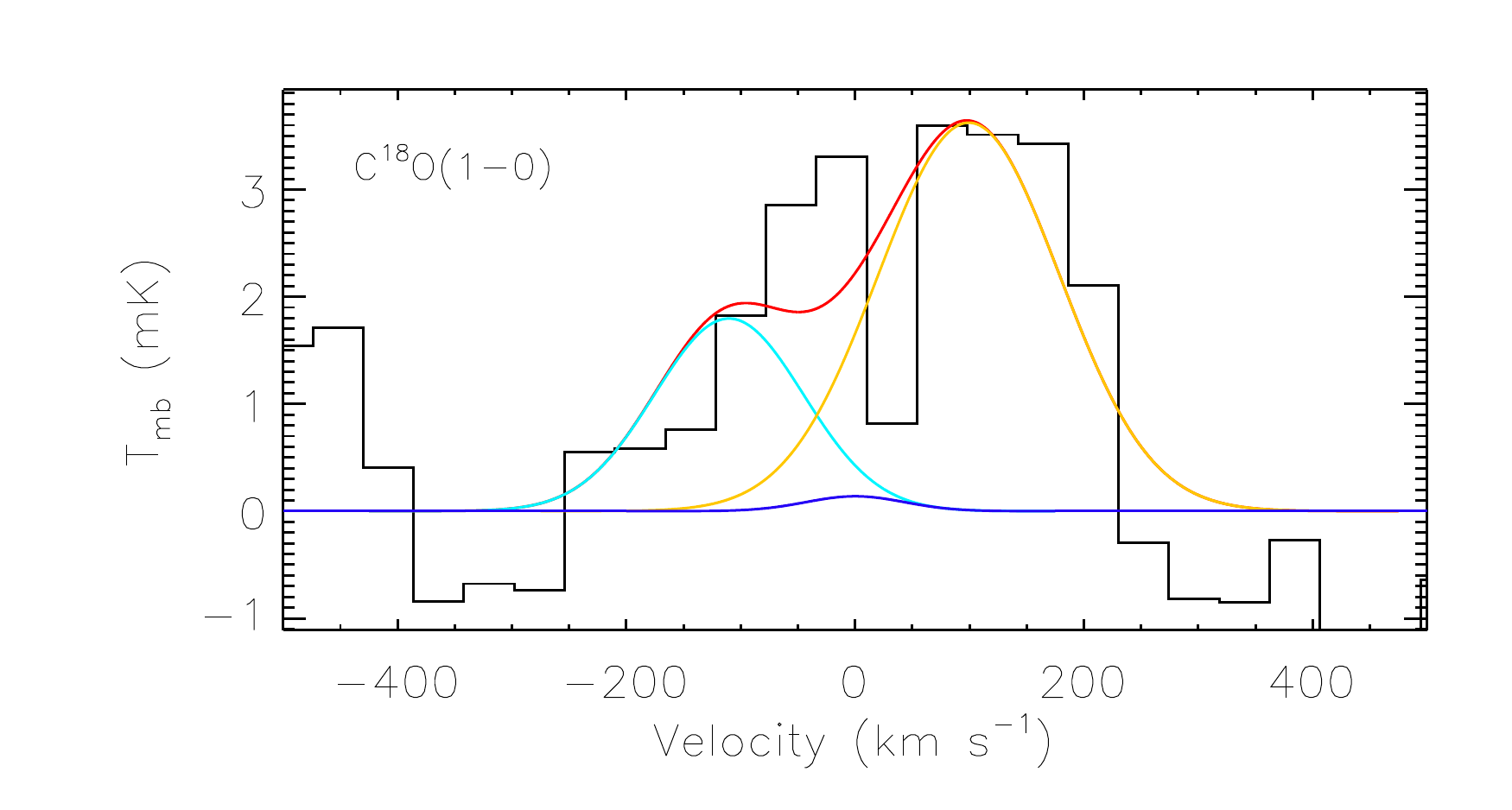}
\includegraphics[width=0.45\textwidth,angle=0,clip,trim=0cm 2.1cm 0cm 1.0cm]{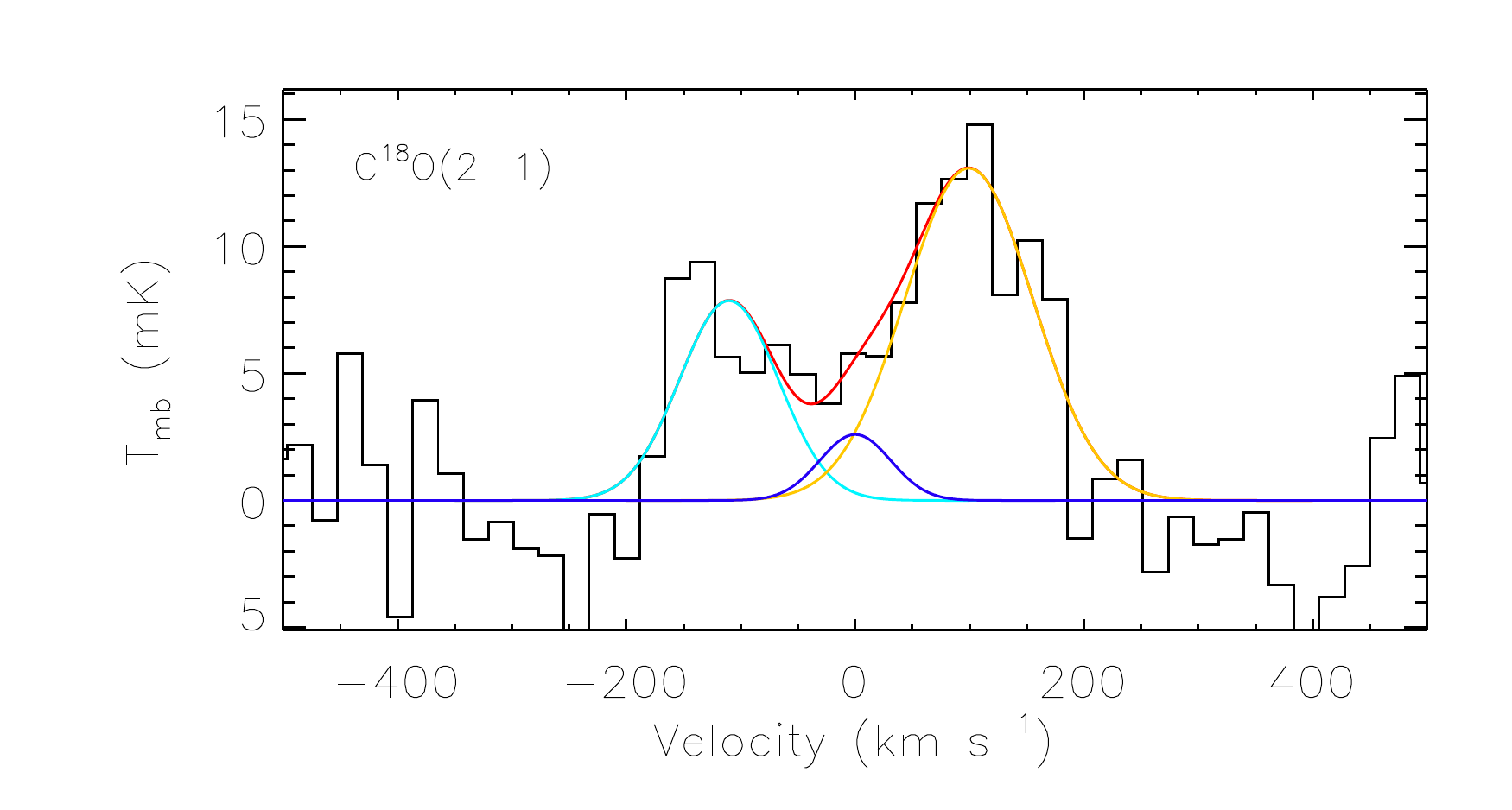}\\
\includegraphics[width=0.45\textwidth,angle=0,clip,trim=0cm 0.0cm 0cm 1.0cm]{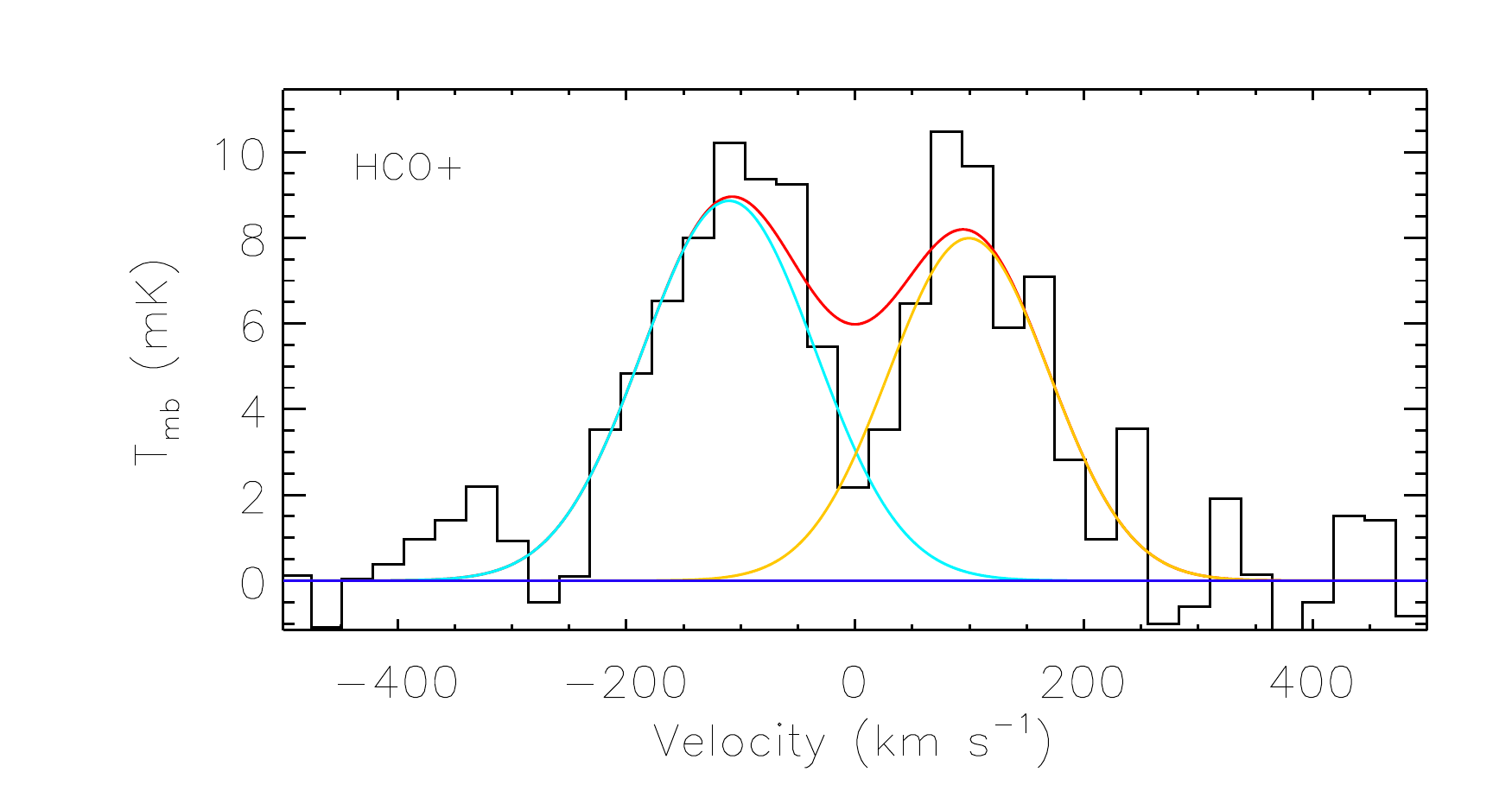}
\includegraphics[width=0.45\textwidth,angle=0,clip,trim=0cm 0.0cm 0cm 1.0cm]{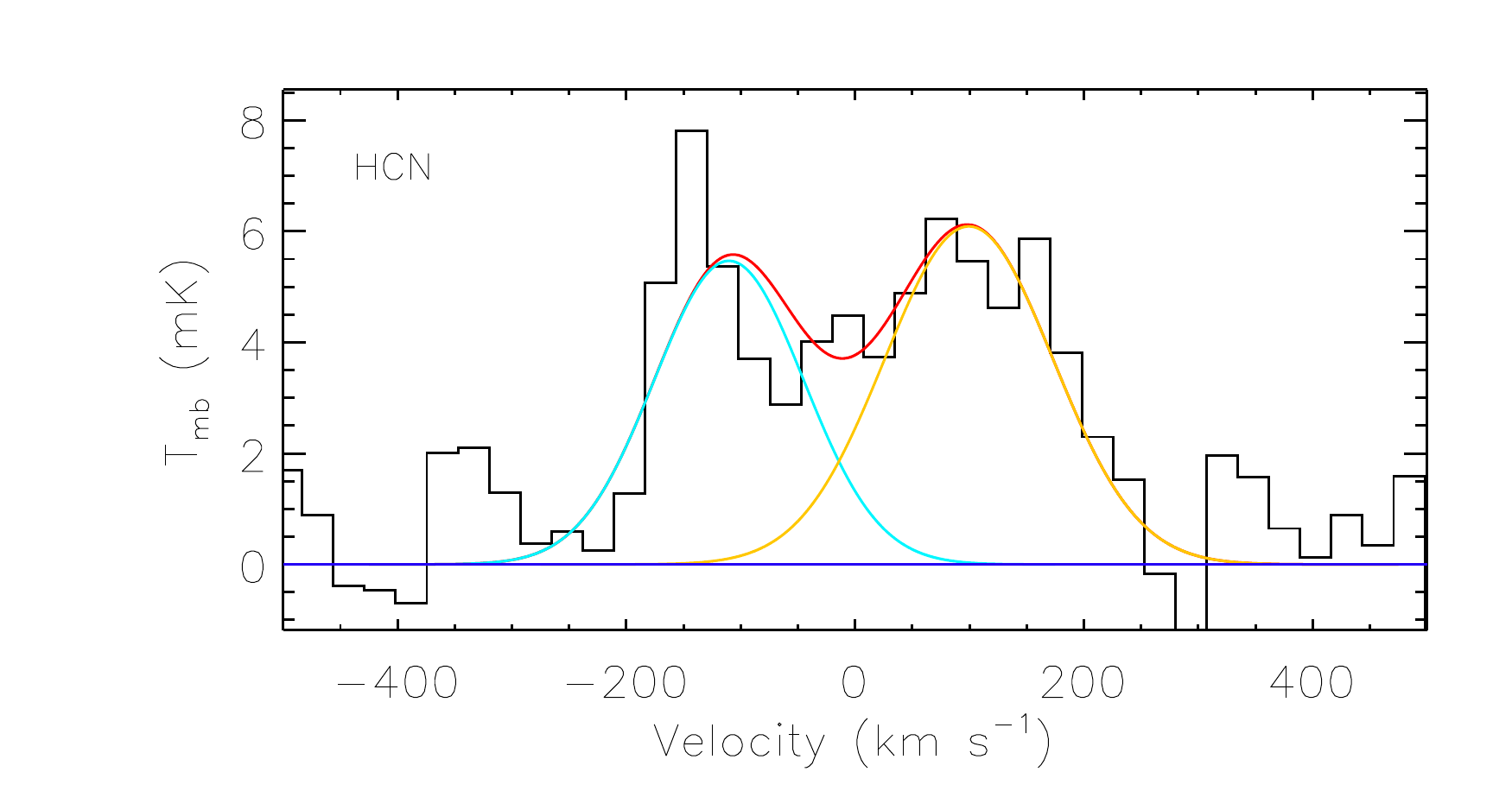}
  \end{center}
\caption{Line profiles in Arp~157 (in units of main beam temperature), overlaid with three fitted gaussians (orange, blue and cyan lines), which together (red line) fit well the observed profile . The gaussain components are constrained to lie at the same central velocities in each plot, based on those found for the observed $^{12}$CO(2-1) line profile in the top right panel. The transition considered is indicated at the top left of each plot. }
\label{lineprof_fitting}
 \end{figure*}

In order to accurately reproduce the observed $^{12}$CO spectra at least three gaussian components are required. The top panel of Figure \ref{lineprof_fitting} shows a 3 gaussian fit to the $^{12}$CO(2-1) spectrum. A central component around the galaxy systemic velocity is required, along with two higher velocity components at $\approx$$\pm$100\kms, with different luminosities. It is possible to fit these three gaussian components to the other well detected lines by varying the line intensities only. The results of this process are shown in the lower panels of Figure \ref{lineprof_fitting}, and in Table \ref{decompfluxtable} (where in converting to beam temperature we have assumed source sizes of 2\farc2 for the red- and blue-shifted components, and 1\farc6 for the central component, based on the channel maps of \citealt{2001ApJ...550..104Y}). 

The component around the kinematic centre of this merger remnant appears to contain little high density molecular gas, as the line profiles of HCO$^+$ and HCN lines can be fitted entirely without emission at these velocities. {The strength of this central component seems to vary systematically as a function of critical density, with molecules like HCN and HCO$^+$ (with critical densities $\approx$700 times higher than CO; \citealt{2007ApJ...656..792P}) not being present at all.} Assuming the ratio of optically thick $^{12}$CO lines to optically thin $^{13}$CO and C$^{18}$O lines vary only with opacity (e.g. no isotope dependant fractionation), it is also clear that this component is significantly {less optically thick} ($^{12}$CO/$^{13}$CO line ratios of $\approx$27) than the higher velocity emission components (typical $^{12}$CO/$^{13}$CO line ratios of $\approx$20). 

One physical interpretation of these line profiles would be if the majority of the dense gas in this system exists in the edge-on starburst ring (inferred to exist from radio continuum and \hi\ absorption observations; \citealt{2003MNRAS.346..424B}), but another molecular component exists near the kinematic centre of this object. The gas in this central component would have to be less dense, and have lower optical depth. The emission from the gas ring would have to be asymmetric, with the red-shifted side {(which corresponds to the eastern side of the starburst ring, further from the centre of the merging galaxy system)} showing enhanced emission. This could arise because of shocks or gas compression as gas flows inwards due to the merger. Alternatively a discrete event like a recent supernova in the ring could be responsible for sweeping up gas and brightening emission from the approaching side of the gas ring. The Br$\gamma$ emission from this source also breaks into three components, corresponding to the edges and centre of the starburst ring see in radio continuum, supporting a ring interpretation \citep{2001A&A...366..439K}.

In a disturbed edge-on system like Arp~157 it is clear that a wide variety of physical models could fit the observed line profiles, not only the one we propose here. Higher resolution interferometric observations of {different lines (e.g. dense gas tracers like HCN and optically thin tracers like C$^{18}$O)} would be required to shed further light on the spatial structure of this object. 

 \begin{table*}
\caption{Integrated beam corrected fluxes from line decompositions. }
\begin{tabular*}{1\textwidth}{@{\extracolsep{\fill}}l r r r r r r}
\hline
Molecule & Peak T$_{\rm b,blue}$ & $\int T_{\rm b,blue}\, \delta V$ & Peak T$_{\rm b,cent}$ & $\int T_{\rm b,cent}\, \delta V$ & Peak T$_{\rm b,red}$ & $\int T_{\rm b,red}\, \delta V$   \\
& (K) & (K \kms) & (K) & (K \kms) & (K) & (K \kms) \\
 (1) & (2) & (3) & (4) &(5)&(6) &(7)\\
 \hline
$ \rm CO(1-0)$ &     19.61 $\pm$      0.10 &   3466.72 $\pm$     20.76 &     22.85 $\pm$      0.28 &   2583.55 $\pm$     34.84 &     24.60 $\pm$      0.10 &   4544.92 $\pm$     21.06\\
$ \rm CO(2-1)$ &      9.46 $\pm$      0.03 &   1688.48 $\pm$      6.17 &      8.26 $\pm$      0.08 &    847.70 $\pm$      9.35 &     12.42 $\pm$      0.03 &   2260.69 $\pm$      6.60\\
$ \rm ^{13}CO(1-0)$ &      0.95 $\pm$      0.12 &    173.29 $\pm$     25.53 &      0.93 $\pm$      0.33 &     97.77 $\pm$     57.01 &      1.58 $\pm$      0.13 &    309.32 $\pm$     31.41\\
$ \rm ^{13}CO(2-1)$ &      0.51 $\pm$      0.06 &     88.98 $\pm$     10.66 &      0.49 $\pm$      0.16 &     47.89 $\pm$     28.70 &      0.79 $\pm$      0.06 &    148.85 $\pm$     14.16\\
$ \rm C^{18}O(1-0)$ &      0.19 $\pm$      0.20 &     30.30 $\pm$     21.43 &      0.03 $\pm$      0.51 &      2.91 $\pm$      2.06 &      0.37 $\pm$      0.23 &     74.66 $\pm$     52.79\\
$ \rm C^{18}O(2-1)$ &      0.20 $\pm$      0.13 &     21.47 $\pm$     12.35 &      0.12 $\pm$      0.26 &      9.45 $\pm$      6.69 &      0.33 $\pm$      0.10 &     46.33 $\pm$     27.01\\
$ \rm HCO+$ &      1.40 $\pm$      0.11 &    264.79 $\pm$     46.64 &      0.00 $\pm$      0.00 &      0.00 $\pm$      0.00 &      1.26 $\pm$      0.13 &    221.91 $\pm$     26.41\\
$ \rm HCN$ &      0.87 $\pm$      0.18 &    143.32 $\pm$     31.97 &      0.00 $\pm$      0.00 &      0.00 $\pm$      0.00 &      0.97 $\pm$      0.15 &    179.85 $\pm$     31.96\\
\hline
\end{tabular*}
\parbox[t]{1 \textwidth}{ \textit{Notes:}  Column 1 indicates the transition considered. Column 2, 3 and 4 contain the integrated flux for the blueshifted, central and redshifted components (respectively), as identified in the line decompositions shown in Figure \protect \ref{lineprof_fitting}. The integrated fluxes have been corrected for beam dilution using Equation \ref{beamdilu}, and assumed source sizes of 2\farc2 for the red- and blue-shifted components, and 1\farc6 for the central component, based on the channel maps of \citealt{2001ApJ...550..104Y}.}
\label{decompfluxtable}
\end{table*}

 \subsection{CO gas properties} 
\label{cotex}
In order to determine the bulk gas properties in the $^{12}$CO emitting gas, which is likely to have a significant optical depth, we used isotopic ratios, and an excitation temperature analysis.

\subsubsection{CO excitation temperature}

The excitation temperature of a spectral line ($T_{\rm ex}$) can be calculated from the below equation.

\begin{equation}
T_{\rm ex}=T_0\left(\ln\left[\left (\frac{T_{\rm b,peak}}{T_0(1-e^{-\tau})}+\frac{1}{e^{T_0/T_{bg}}-1}\right)^{-1}+1\right]\right)^{-1}
\label{texeqn}
\end{equation}
\noindent where $T_0 = h\nu/k_b$ [=5.53 K for $^{12}$CO(1-0)], $\nu$ is the frequency of the transition, $h$ is Planck's constant, $k_b$ is Boltzmann's constant, $T_{\rm b,peak}$ is the peak beam temperature in Kelvin, $\tau$ is the optical depth of the line in question (here the $^{12}$CO(1-0) line), and $T_{\rm bg}$ is the background radiation field temperature (usually taken as the cosmic microwave background temperature of 2.725K). 

The optical depth ($\tau$) of the $^{12}$CO(1-0) transition can be estimated from the ratio of the integrated intensities of $^{12}$CO(1-0) and $^{13}$CO(1-0), assuming that $^{13}$CO(1-0) is optically thin, and that we know the $^{13}$CO abundance with respect to $^{12}$CO. We do not detect sufficient isotopologues to constrain these abundance values directly; however the ISM in this object has approximately solar metallicity \citep{1991ApJ...381..409S}.  Thus we assume the standard solar neighbourhood abundance of  $^{12}$C/$^{13}$C =70  \citep[and $^{12}$C/C$^{18}$ =250;][]{1994ARA&A..32..191W}. {This value lies in the middle of the range found for extragalactic objects ($^{12}$C/$^{13}$C$\approx$40-130; e.g. \citealt{1993A&A...274..730H,2000A&A...358..433M,2010A&A...522A..62M}).}
Using these assumptions we find that the optical depth of the $^{12}$CO lines (for both the source average and the different kinematic components) to be $\approx$2.8-4.5  (see Table \ref{source_split_table}).

{Assuming that $^{13}$CO emission arises from regions with the same excitation temperature as $^{12}$CO(1-0), and additionally that the $^{12}$CO is optically thick (which we showed seems to be the case above), the expression for $\tau$($^{13}$CO) has the following very simple form.} 

\begin{equation}
\tau(^{13}\rm CO) = \ln\left[\left(1-\frac{T_{\rm b}[\rm ^{13}CO(1-0)]}{T_{\rm b}[\rm ^{12}CO(1-0)]} \right)^{-1}\right]
\end{equation}
\noindent where $\tau(^{13}\rm CO)$ is the optical depth of $^{13}$CO, and $T_{\rm b}[X]$ is the observed main beam brightness temperature for transition X.  Here we have neglected a correction factor, which is equal to unity if $T_{\rm ex}$ is equal for $^{12}$CO and $^{13}$CO and $T_{\rm ex}>>T_{\rm bg}$ (as we show above $T_{\rm ex}$ for all components is $>$15K, validating this assumption).
Using the values from Table \ref{obstable} and \ref{decompfluxtable} we find a negligible $\tau(^{13}\rm CO)\ltsimeq$0.07 in all cases (Table \ref{source_split_table}). 
{The same is found if we relax the assumption that $^{12}$CO is optically thick, and instead assume that $\tau$($^{12}$CO)~=~[$^{12}$CO/$^{13}$CO]/$\tau$($^{13}$CO) (where we assume the [[$^{12}$CO/[$^{13}$CO] ratio is tracing the abundance of $^{12}$CO with respect to $^{13}$CO).}
This suggests we are justified in making the assumption that $^{13}$CO is optically thin in this source.

Combining these estimates of the optical depth with the $T_{\rm b,peak}$ source average values from Table \ref{obstable} we find an excitation temperature of 14.8 K for $^{12}$CO(1-0) in Arp~157 (and a $^{12}$CO(2-1) excitation temperature of 8.9K).  
In the individual kinematic components (identified in Section \ref{linesfound}) using the values in Table \ref{decompfluxtable} we find excitation temperatures of between 23K and 28K for $^{12}$CO(1-0), and between 13-17.5 K for $^{12}$CO(2-1), as listed in Table \ref{source_split_table}.
{The lower excitation temperatures found for the CO(2-1) transition imply that this line may come from regions deeper within the molecular clouds in this object, where densities are higher and temperatures lower. }

\subsubsection{Column densities and masses}
\label{cotex_column}
Using the excitation temperatures calculated in the previous section it is possible to directly calculate the column density for a given molecule, in each kinematic component using the following formalism (see appendix B in \citealt{2006ApJS..164..450M} for more details).

\begin{equation}
{\rm N} =  \frac{8 \pi k_b \nu^2}{h c^3 A_{\rm ul} g_{\rm u}} \eta(T_{ex},T_{bg},\tau) Z(T_{\rm ex}) W e^{Eu/k_bT_{\rm ex}}
\end{equation}

\noindent where N is the total column density of a given species, W is $\int T_{\rm b}\, \delta V$ for that transition, A$_{\rm ul}$ is the Einstein A coefficient for the upper level, $g_u$ is the degeneracy of the upper level, $Z(T_{\rm ex})$ is the partition function at temperature $T_{\rm ex}$, calculated as described below, and $\eta(T_{ex},T_{bg},\tau)$  is a correction term that takes into account the effect of optical depth, and the background temperature of the region:

\begin{eqnarray}
\eta(T_{ex},T_{bg},\tau)=\left( \frac{\tau}{1-e^{-\tau}}\right)\left(1-\frac{J_{\nu}(T_{bg})}{J_{\nu}(T_{ex})} \right)^{-1}\\
J_{\nu}(T)=\frac{h\nu}{k_b}\left( e^{h\nu/k_bT} -1\right)^{-1}
\end{eqnarray}
\noindent where the symbols are as previously defined in this section. 
 As we showed in Section \ref{cotex}, $T_{\rm ex}$ is large when compared to the background temperature in this source, and $^{13}$CO is optically thin. Hence in this section we can assume $\eta(T_{ex},T_{bg},\tau)$=1. 

The partition function for simple molecules like carbon monoxide can be defined as:

\begin{equation}
Z = \sum_{i=0}^{\infty}g_i e^{-E_i/k_b T_{\rm ex}}
\label{partfunc}
\end{equation}
where $g_i$ and $E_i$ are the degeneracy and upper state energy of the $i$'th level of a given molecule. 

We here use these equations to estimate the total column density of $^{13}$CO for the three kinematic components identified above, retrieving the required physical constants from the Cologne Database for Molecular Spectroscopy\footnote{http://www.astro.uni-koeln.de/cdms/} \citep[CDMS;][]{2001A&A...370L..49M,2005JMoSt.742..215M}. We calculate the total column densities using the $^{13}$CO(1-0) values presented for the different kinematic components in Table \ref{decompfluxtable}, and use the excitation temperatures derived in Section \ref{cotex}.
Note that here we use the excitation temperature derived for $^{12}$CO, and are thus explicitly assuming that $^{12}$CO and $^{13}$CO arise from regions with the same excitation properties (which may not be valid). 
The derived total column densities are shown in Table \ref{source_split_table}. 

 Column densities in extragalactic sources are an average over a complex ensemble of clouds of different densities (as well as empty regions), and hence it often makes sense to compare the total number of molecules of a given species present in the source, rather than column densities. Table \ref{source_split_table} shows these values, calculated assuming the source sizes described above. {The values for the individual kinematic components when summed slightly over predict the source average value. This difference is likely not significant within our errors.}

 From these values it {is possible to roughly estimate} the total mass of H$_2$ present in the system, if one assumes a value for its fractional abundance with respect to $^{13}$CO. This value is not known in Arp~157, but here we use a value of 7$\times$10$^{-5}$ (in line with our assumed $^{12}$CO/$^{13}$CO abundance of 70, and a  Galactic H$_2$ to $^{12}$CO abundance ratio of 10$^{-4}$; \citealt{1988ApJ...334..771V}). We caution that this value is uncertain by \textit{at least} a factor of 2 \citep[e.g.][]{1992IAUS..150..285V}. The sum over the three kinematic components would imply a total H$_2$ mass of 1.34$\times$10$^9$ M$_{\odot}$ in this object, consistent within the uncertainties with the estimate of $\approx1.9\times10^{9}$ M$_{\odot}$ made by \cite{2001ApJ...550..104Y}, based on their interferometric $^{12}$CO flux measurement and assuming a standard galactic $X_{CO}$ factor.

Overall we conclude that the different kinematic components of this source show systematic differences in properties. The redshifted part of the putative ring is hotter, more optically thick, and contains somewhat more material than the blue-shifted side.  The component around the kinematic centre of the system is less dense, and more optically thin than the "ring" components. If these differences arise as a result of the ongoing merger, or a consequence of extreme star-formation activity will only become clear by obtaining {higher resolution resolved data of different molecular species}.

 \begin{table}
\caption{Properties of the gas in different kinematic source components.}
\begin{tabular*}{0.47\textwidth}{@{\extracolsep{\fill}}l r r r r r}
\hline
Property & Total & Blue & Central & Red & Unit \\
(1) & (2) & (3) & (4) & (5) & (6) \\ 
 \hline
$T_{\rm ex}$[$^{12}$CO(1-0)] & 14.3 & 23.1 & 26.4 & 28.1 & K \\
$T_{\rm ex}$[$^{12}$CO(2-1)] & 8.9 & 14.5 & 13.2 & 17.56 & K \\
$\tau$[$^{12}$CO(1-0)] & 4.37 & 3.39 & 2.84 & 4.49 & -\\
$\tau$[$^{12}$CO(2-1)] & 4.08 & 3.77 & 4.15 & 4.45 & -\\
$\tau$[$^{13}$CO(1-0)] & 0.06 & 0.04 & 0.04 & 0.07 & -\\
$\tau$[$^{13}$CO(2-1)] & 0.06 & 0.05 & 0.06 & 0.07 & -\\
N($^{13}$CO) & 2.14 & 2.47 & 1.54 & 5.12 & 10$^{17}$ cm$^{-2}$\\
n$_{\rm mol}(^{13}$CO) & 3.51 & 1.64 & 0.54 & 3.40  & 10$^{61}$ \\
M$_{\rm H_2}{\dagger}$& 8.5 & 3.9 & 1.3 & 8.1 & 10$^{8}$ M$_{\odot}$ \\
\hline
\end{tabular*}
\parbox[t]{0.47 \textwidth}{ \textit{Notes:} This table contains a list of the derived properties of Arp~157 as a whole, and for each kinematic component (as identified in \ref{linesfound}).The property in question is listed in Column 1, the values in Columns 2-5, for the whole source, and the blueshifted, central and redshifted components, respectively. The units are shown in Column 6. Here $T_{\rm ex}$ is the excitation temperature, $\tau$(X) is the optical depth of transition X, N($^{13}$CO) is the total column density of $^{13}$CO, N$_{\rm mol}(^{13}$CO) is the total number of $^{13}$CO molecules, and M$_{\rm H_2}$ is the H$_2$ mass calculated assuming a $^{13}$CO-to-H$_2$ abundance of 7$\times$10$^{-5}$.}
\label{source_split_table}
\end{table}

\subsection{Rotation Diagrams}  
 \label{rotdiag_sec}

In order to estimate the column density of other molecules we detect here, we can use rotation diagrams. 
From the observed beam corrected brightness temperature of the
measured lines of any molecule one can directly estimate the column density of these molecules in the upper level of the observed transition, assuming the emission is optically thin. The column density in the upper level of the observed transition (N$_u/$g$_u$; where g$_u$ is the degeneracy of the state) is calculated as in Equation \ref{eqnnmugmu}  (see \citealt{1999ApJ...517..209G} and Appendix  B1 of \citealt{2006ApJS..164..450M} for a detailed discussion of this technique).

\begin{equation}
\frac{N_u}{g_u} =  \eta(T_{ex},T_{bg},\tau) \left(\frac{8 \pi k_b \nu^2}{hc^3 A_u g_u}\right) \left(\frac{\int T_b\, \delta V}{\rm K\,km s^{-1}}\right) \\
\label{eqnnmugmu}
\end{equation}
\noindent where the symbols have the same meanings as above, with $\nu$ is the frequency of the transition, and $\int T_b\, \delta V$ is the integrated intensity of the beam corrected brightness temperature, in units of K \kms, as tabulated in Table \ref{obstable}. $\eta(T_{ex},T_{bg},\tau)$, as discussed above, is a correction term that takes into account the effect of optical depth, and the background temperature of the region. In this work we assume $\eta(T_{ex},T_{bg},\tau)$=1, which is only valid if $T_{ex}>>T_{bg}$, and if the transition in question is optically thin. The assumption that the emission detected arises from optically thin regions may be violated for some of the high critical density molecules detected in this work (where $\tau$ is likely to be $>1$). As the comparison samples we use in this work all assume that all the gas is optically thin we continue with this assumption from here on, but caution that if optical depths are significant in some molecules, our derived column densities are lower limits.  

In order to convert the column densities
in the observed states to a total (source average) column density for a given molecule, one can use a rotation diagram, as long as more than one transition has been detected. See \cite{1999ApJ...517..209G,2006ApJS..164..450M} for full details of this method. 
The derived column densities depend on the partition function, evaluated at the derived rotation temperature. In molecules with complicated structures, where Equation \ref{partfunc} is not valid (e.g. CH$_3$OH) we use linear interpolations based on the tabulated partition functions available from CDMS.

With the observations we possess here we are usually only able to derive a single rotational temperature for each gas tracer. In reality several gas components with different rotation temperatures are often required to fit observational data \citep[e.g.][]{2009ApJ...707..126B}.The colder components of the ISM that emit in low $J$ lines usually dominate the total column density estimate, suggesting our derived column densities should be reasonably robust, and at worst can be treated as lower limits. \cite{2013MNRAS.433.1659D} found adding higher-J lines increase total column density estimates by less than $<$10\% on average (for their sample of early-type galaxies).

Some of the lines we detect in this survey are blended (e.g. methanol, C$_2$H and CN). In order to create rotation diagrams and calculate the total column density for these molecules we use a statistical decomposition (from Appendix B2 in \citealt{2006ApJS..164..450M}), which is based on the relative ratios of the Einstein coefficients (once again retrieved from the CDMS). For methanol, where the blending is not due to hyperfine structure one must assume an internal rotation temperature in order to split the levels. Here (as in \citealt{2006ApJS..164..450M}) we use a value that minimises the difference between the slope derived from a single set of blended lines, and the slope derived from both sets of lines. This statistical method for splitting blended lines should be treated with caution, as the error on the derived quantities may be much higher than reflected in the formal error bars (especially for methanol, where the few transitions observed represent only a small fraction of the emitting column density).

\begin{figure}
\begin{center}
\includegraphics[width=0.43\textwidth,angle=0,clip,trim=1.0cm 0cm 0.8cm 1.0cm]{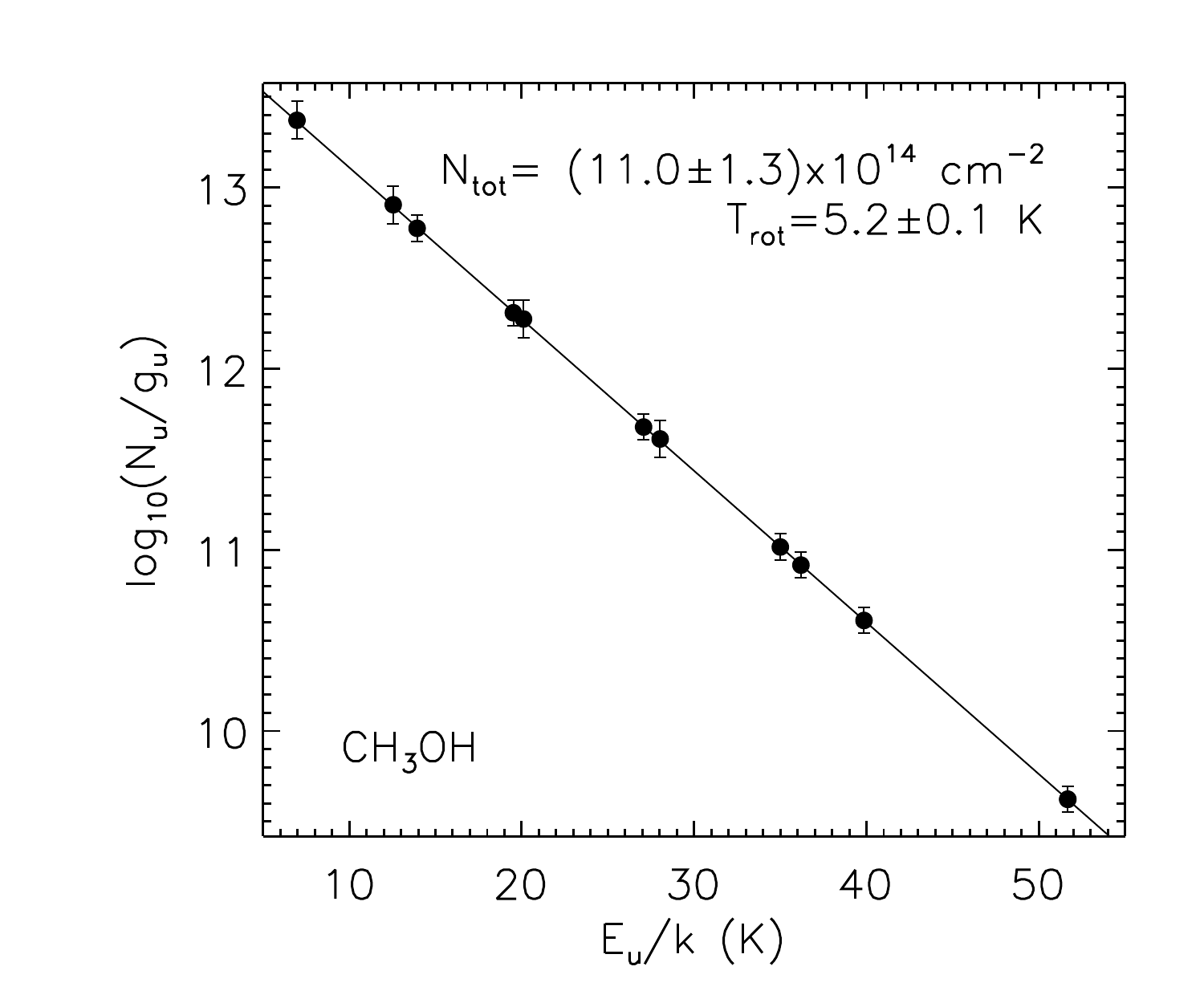}
\includegraphics[width=0.43\textwidth,angle=0,clip,trim=1.0cm 0cm 0.8cm 1.0cm]{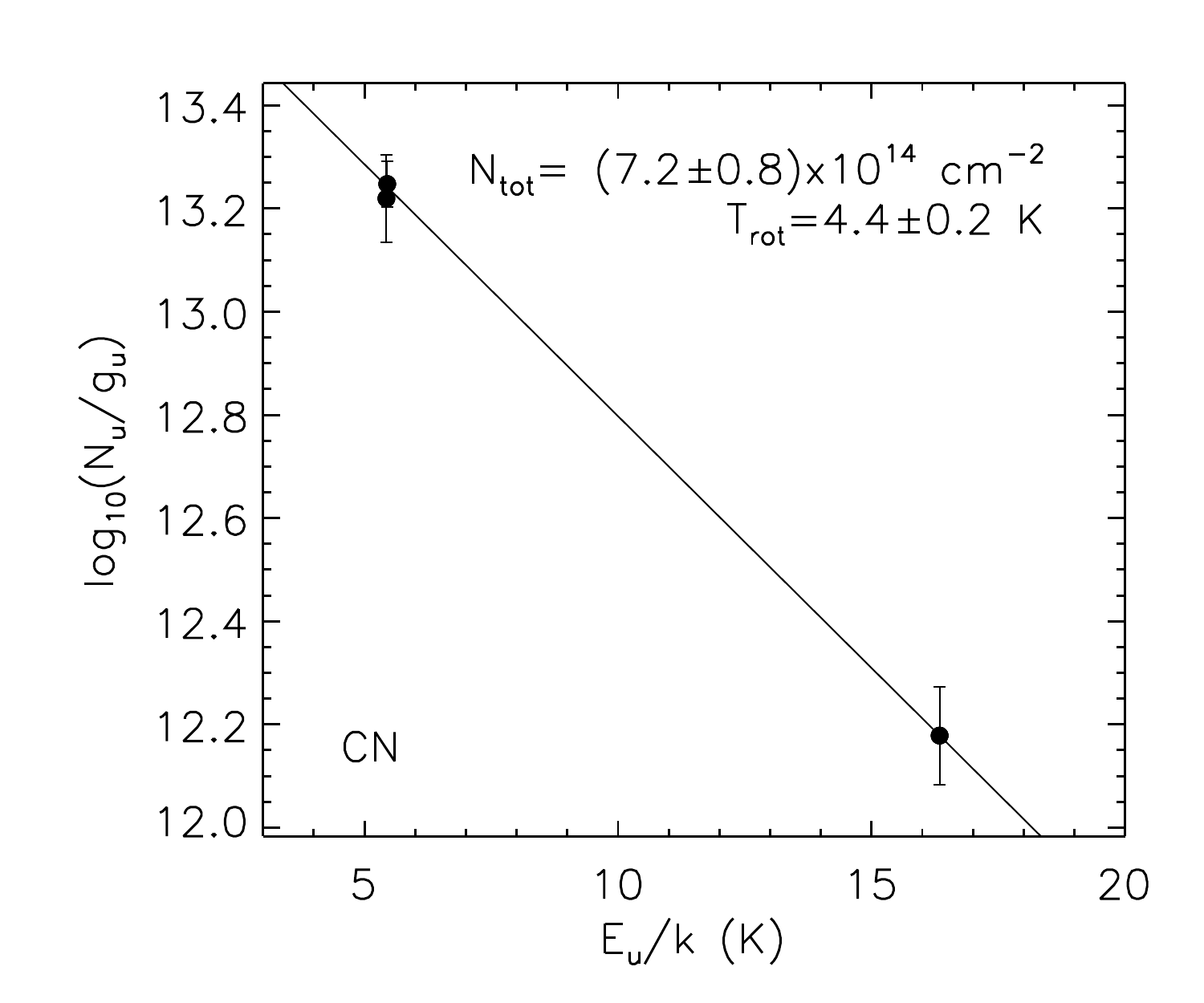}
 \end{center}
 \caption{Rotation diagrams for the chemical species with more than two fitted data points. The energy of the upper level is plotted against the derived column density in the upper level for each observed transition (black points), as tabulated in Table \ref{coldenstable}. These data points are fitted with a line and the best fit, corresponding to the rotation temperature indicated in the legend, is displayed (solid line). We split methanol and CN blends statistically, as described in Section \ref{rotdiag_sec}.}
 \label{rotdiags}
 \end{figure}

Figure \ref{rotdiags} shows the rotation diagrams constructed in this way, for each molecule where we have more than two independent fitted points. {The rotation diagrams for CO isotopes and CS are not shown here, as it is always possible to fit a line exactly between two points, and thus showing these diagrams gives no information about the quality of our single temperature fit.} In order to estimate the column density of the cases where only one transition was detected we used the measured mean T$_{\rm rot}$ of the other detected species (T$_{\rm rot}$=5.7K).  Table \ref{coldenstable} shows the derived total column densities, total number of molecules, and rotation temperatures for the detected species. In Table \ref{coldenstable} we include in the stated uncertainties the effect of varying the assumed temperature by $\pm$2 K. {This value is chosen because it covers the full range of temperatures found for molecules where we have the information to derive rotation diagrams.}
In Table \ref{coldenstable} we also present results for the  $^{13}$CO decomposed line profiles, fitting rotation diagrams for each component separately.

\begin{table*}
\caption{Total column densities and rotation temperatures derived from rotational diagrams. }
\begin{tabular*}{0.8\textwidth}{@{\extracolsep{\fill}}l r r r r r}
\hline
Molecule & Gradient & Intercept & Temp & Total Col. Dens. & Num Mols. \\
& (K$^{-1}$) & log$_{10}$(cm $^{-2}$)  & (K) & (10$^{14}$ cm $^{-2}$)  & (10$^{59}$)\\
 (1) & (2) & (3) & (4) & (5) & (6) \\
 \hline
C$_2$H& - & - & 5.7$^*$ & 15.4 $\pm$ 3.8 & 2.53 $\pm$ 0.62\\
HNCO& - & - & 5.7$^*$ & 1.5 $\pm$ 0.5 & 0.24 $\pm$ 0.08\\
HCN& - & - & 5.7$^*$ & 2.6 $\pm$ 0.1 & 0.43 $\pm$ 0.02\\
HCO$^+$& - & - & 5.7$^*$ & 1.5 $\pm$ 0.1 & 0.24 $\pm$ 0.01\\
HNC& - & - & 5.7$^*$ & 0.8 $\pm$ 0.1 & 0.13 $\pm$ 0.01\\
CS &-0.09 & 13.88 & 4.6 $\pm$ 0.41 & 3.2 $\pm$ 1.0 & 0.53 $\pm$ 0.16\\
CH$_3$OH &-0.08 & 13.95 & 5.2 $\pm$ 0.10 & 11.0 $\pm$ 1.3 & 1.81 $\pm$ 0.21\\
C$^{18}$O &-0.06 & 15.87 & 6.9 $\pm$ 1.49 & 211.6$^{+148.}_{-96.}$ & 34.79 $\pm$ 20.04\\
$^{13}$CO &-0.08 & 16.54 & 5.3 $\pm$ 0.16 & 1648.4 $\pm$ 127.7 & 271.03 $\pm$ 21.00\\
\hspace{8pt}$\rightarrow$\hspace{8pt}blue &-0.08 & 16.48 & 5.2 $\pm$ 0.48 & 1393.54 $\pm$ 413.55 & 92.63 $\pm$ 27.49\\
\hspace{8pt}$\rightarrow$\hspace{8pt}cent &-0.09 & 16.24 & 5.0 $\pm$ 2.01 & 789.98$^{+1759.}_{-571.}$ & 27.77 $\pm$ 40.96\\
\hspace{8pt}$\rightarrow$\hspace{8pt}red &-0.09 & 16.74 & 5.0 $\pm$ 0.33 & 2504.13 $\pm$ 523.22 & 166.46 $\pm$ 34.78\\
CN &-0.10 & 13.77 & 4.4 $\pm$ 0.18 & 7.2 $\pm$ 0.8 & 1.18 $\pm$ 0.13\\
H$_2$CO& - & - & 5.7$^*$ & 0.8 $\pm$ 0.1 & 0.12 $\pm$ 0.01\\
\hline
\end{tabular*}
\parbox[t]{0.8 \textwidth}{ \textit{Notes:}  Column 1 indicates the transition considered. In the case of $^{13}$CO we present results for the decomposed line profiles, and this is also indicated in Column 1. Column 2 and 3 contain the gradient and intercept of the best fit line on the rotation diagrams displayed in Figure \ref{rotdiags}. Column 4 shows the derived rotation temperature. Where the value is marked with a star the temperature is assumed to be 5.7$\pm$2 K. Column 5 is the derived total source-averaged column density of this species. Where errors are notably asymmetric we show both an upper and lower bound. Column 6 is the derived total number of molecules of this species}
\label{coldenstable}
\end{table*}

The derived gas kinetic temperatures presented in Table \ref{coldenstable} vary little, with the dense gas tracers where we detect two or more transitions (CS, CN, CH$_3$OH) having rotational temperatures of $\approx$4.5-5.5K, {and CO isotopes having temperatures of} $\approx$5.3-7K. This analysis assumes all transitions are optically thin, which makes it somewhat surprising that between different tracers the derived temperatures vary little.

These temperatures are lower than the estimates of the CO excitation temperatures, as derived above.
The rotation diagram method assumes that the lines are optically thin and that for each transition the emission comes from the same region, while the excitation temperature analysis assumes that the molecules have the same T$_{\rm ex}$. Taking $^{13}$CO as a test case, where we have already ascertained that the emission is optically thin, we propose that this difference in temperatures can be explained by a change in the emitting area for the higher $J$ emitting gas. If the $^{13}$CO(2-1) emission in this source comes from an area $\approx$50\% smaller than the $^{13}$CO(1-0) emission we would obtain a T$_{\rm rot}$ of 15K, similar to the source-average T$_{\rm ex}$ derived above. We only have the isotopic data required to demonstrate this directly for transitions of CO, but this illustrative case warns us that the derived T$_{\rm rot}$ for other species could be underestimated as well, if the emitting regions for higher excitation lines are smaller.
Resolved data for these line species would be required in order to determine the true source sizes for all emitting species and transitions. Such data does not exist for any of the molecular species under study here. The total number of molecules, and the molecular source average column densities in Table \ref{coldenstable} all depend on the assumed temperature, and we caution that the uncertainty in our derived values due to a changing source size is almost certainly larger than the formal error which is quoted here. Ratios of the total number of molecules or column densities, however, vary much less strongly, by $\ltsimeq$0.2 dex for any rotation temperature within the range 3-50K (bracketing the measured excitation temperatures above). We hence will use such ratios in later sections.

As discussed above, Table \ref{coldenstable} also includes rotation temperature, column density and estimates of the total number of molecules, derived from the $^{13}$CO decomposed line profiles. We only present these for $^{13}$CO as this is the strongest line (other than the optically thick $^{12}$CO), and thus the additional errors induced by the decomposition are not dominant. Because of the lower rotation temperature, the total column density and estimate of the number of molecules are somewhat smaller than those estimated using an excitation analysis in Section \ref{cotex}. Despite this, we find that the rotation temperatures derived for the different kinematic components are very similar to the source average rotation temperature. The central component has the lowest column density, as predicted in previous sections. {The total column densities derived from these rotation diagrams predict a similar total number of  $^{13}$CO molecules as found in our source average.} We thus continue from this point by using source average values, which we can estimate for all of the detected transitions.

The column densities and temperatures derived for molecules other than CO in this source are similar to those found for nearby spiral and starburst galaxies \cite[e.g.][]{2009ApJ...707..126B}, suggesting that the ongoing merger in this galaxy is not significantly impacting the physical conditions (e.g. temperature, density) within the low $J$-line emitting dense gas reservoir.

 \subsection{Chemical Abundances}
 \label{chem_abund}

Using the source average number of molecules we derived in Section \ref{rotdiag_sec}, we can attempt to determine in more detail what sorts of conditions drive the chemistry in this object.

{The ratio of the total number of molecules of a given species present in a source (or equivalently ratios of the column densities of these molecules) give us an estimate of the abundance ratios in this source (assuming the emission is optically thin, as discussed above). }
As is typically done in the literature (e.g. \citealt{2006ApJS..164..450M,2011A&A...535A..84A,2013A&A...549A..39A}) we here assume that the optical depth of all of these species is low, and our ratios do relate to molecular abundances, and compare and contrast the determined abundances ratios with those found in other well studied sources. 

When conducting this exercise it is usual to normalise ratios with an optically thin tracer, that does not suffer from isotope dependant fractionation (e.g. C$^{34}$S in \citealt{2011A&A...535A..84A}).  In this relatively shallow survey we do not detect a suitable tracer, so have taken the decision to normalise the ratios with respect to the values derived for CS, which in the worst case could be moderately saturated \citep[e.g.][]{1989A&A...223...79M}. {$^{32}$S/$^{34}$S ratios have been shown to be fairly constant in nearby galaxies \citep{2009ApJ...694..610M}, suggesting variations in intrinsic abundance ratio should not introduce a strong bias between objects here.}

\subsubsection{Comparison with other galaxies}
\label{chem_abund_gals}

\begin{figure*}
\begin{center}
\includegraphics[width=0.95\textwidth,angle=0,clip,trim=0.0cm 0.0cm 0cm 0cm]{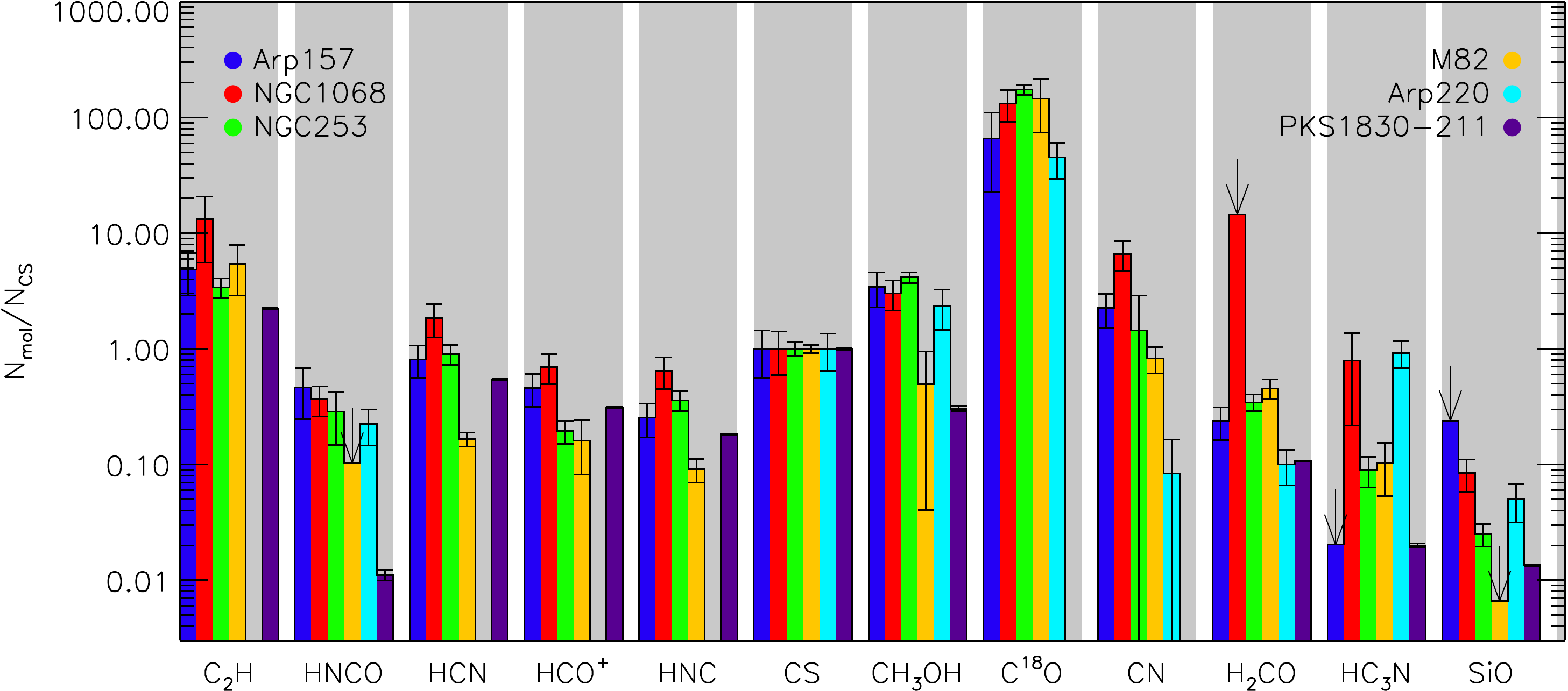}
 \end{center}
 \caption{Comparison of the fractional abundances in Arp~157 (blue), in the AGN NGC~1068 (red), and in two starbust galaxies NGC~253 (green) and M82 (orange). The Y-axis shows the total number of  molecules of a given species, normalised to CS. Grey shaded regions act as a guide to the eye for each set of molecules. Error bars indicate the formal error in the ratio only. Arrows indicate upper limits. The CN abundance for NGC~253 is estimated assuming CN is 1.6$\times$ brighter than HCN in this source, as found by \protect \cite{1995A&A...295..571H}.}
 \label{abundfig}
 \end{figure*}

\begin{figure*}
\begin{center}
\includegraphics[width=0.95\textwidth,angle=0,clip,trim=0.0cm 0.0cm 0cm 0cm]{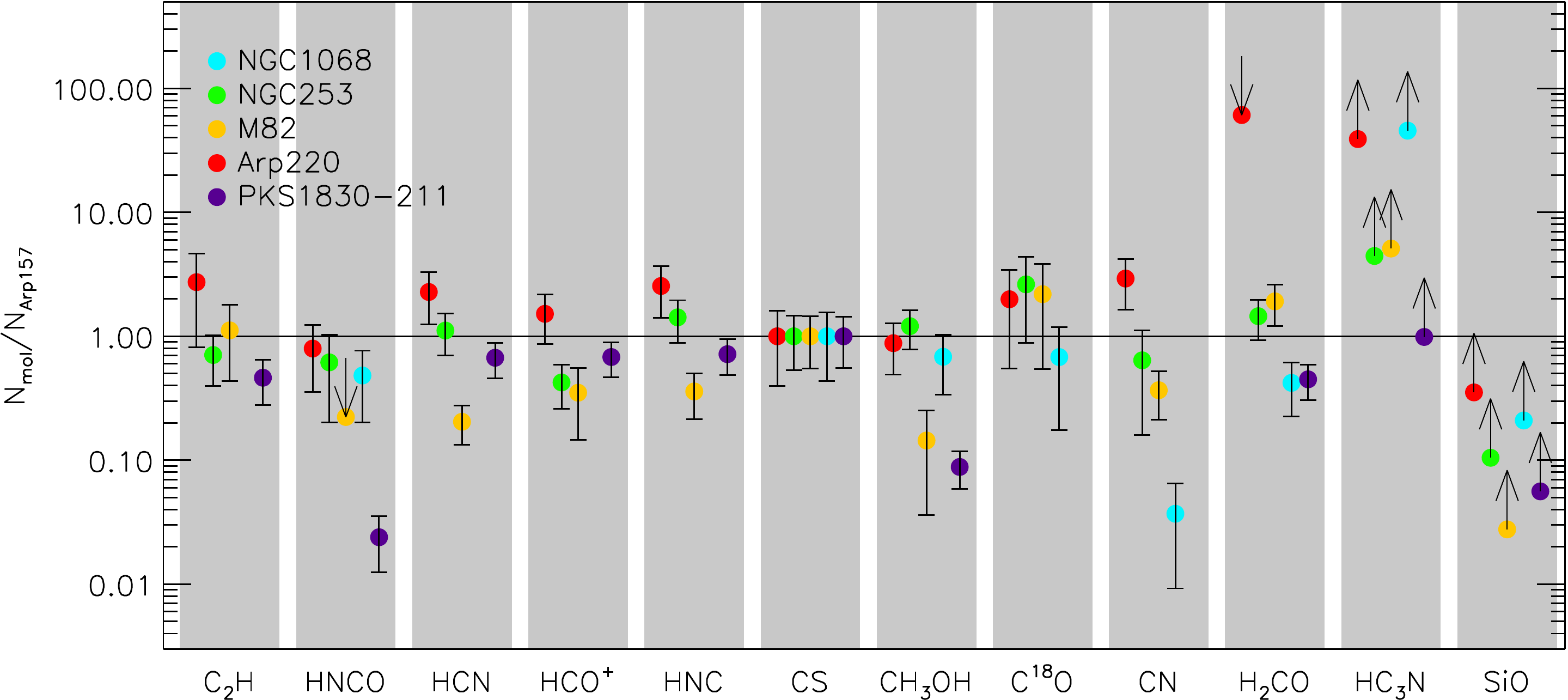}
 \end{center}
 \caption{Comparison of the fractional molecular abundances (with respect to CS) in the AGN NGC~1068 (red), and in two starbust galaxies NGC~253 (green) and M82 (orange), normalised to the abundance of that molecule in Arp~157 (blue). Molecules which were not detected in Arp~157 are included as upper limits. All other details are as in Figure \protect \ref{abundfig}.}
 \label{abundfig_diff}
 \end{figure*}

 In Figure \ref{abundfig} we show our derived abundances (with respect to CS) for the molecules detected in Arp~157.
 We also show the same ratios for a sample of nearby galaxies with line surveys available in the literature. This includes the nearby Seyfert 2 galaxy NGC~1068 \citep{2013A&A...549A..39A}, {ULIRG Arp~220 \citep{2011A&A...527A..36M}} and the starbursts M82 \citep{2011A&A...535A..84A} and NGC~253 \citep{2006ApJS..164..450M}. {We also compare with the molecules detected (in absorption) in an intermediate redshift spiral ($z$=0.89) which is in the line of sight of quasar PKS 1830-211 \citep{2006A&A...458..417M,2011A&A...535A.103M}}. Histogram bars with errors show the abundances and their formal error, while a downwards arrow indicates an upper limit. Figure \ref{abundfig_diff} shows the fractional molecular abundances of these sources (with respect to CS), normalised by that found in this paper for Arp~157. These allow us to more formally identify galaxies with similar abundances, and show easily the relative trends. The standard deviation of the points around the 1:1 line in this figure, {and the reduced $\chi^2$ values (which have the advantage of including the errors)} provide quantitative ways to estimate which galaxy has the most similar abundance to Arp~157, and these values are listed in Table \ref{goodfittable}. 

M82 has a relatively old starburst at its core (when compared with NGC~253), with an average stellar population age of $\approx$10 Myr \citep{2009ApJ...701.1015K}. The strong UV fields this creates cause PDRs to dominate its chemistry \citep{2011A&A...535A..84A}. 
Figure \ref{abundfig} shows that M82 has lower HNCO, HCN, HCO+, HNC, CH$_3$OH and CN abundances than Arp~157, and enhanced C$^{18}$O and H$_2$CO abundances. 
As is clear in Figure \ref{abundfig_diff}, and the intermediate standard deviation value in Table \ref{goodfittable}, it seems that a pure PDR like that present in M82 is not the best template to explain the molecular abundances we observe in Arp~157. 

{Arp~220 \cite{2011A&A...527A..36M} is the most similar object to Arp~157 in our comparison sample. It is an ultra luminous infrared galaxy (ULIRG), with an even higher infrared luminosity than Arp~157. 
It is also an advanced merger system, with large tidal tails, and two visible nuclei. Although X-ray observations of Arp~220 have suggested that both of these nuclei may be Compton-thick AGN \citep{2013arXiv1303.2630P}, the interstellar radiation field in this object is thought to be dominated by a large ongoing starburst \citep{2011A&A...527A..36M}.
 The line survey of this object was conducted with the Sub-millimetre Array (SMA),  at 1.3mm. Unfortunately this means the number of species which overlap with our own line survey is low. The main isotope of CS was also not contained within the survey, and here we have assumed a CS abundance using the C$^{34}$S abundance of \cite{2011A&A...527A..36M}, and a $^{32}$S/$^{34}$S abundance of 8$\pm$2 \citep[e.g.][]{2005ApJ...620..210M,2009ApJ...694..610M}. Taking this estimate, Figure \ref{abundfig} suggests that Arp~220 has HNCO, C$^{18}$O and methanol abundances similar to Arp~157; however CN is less abundant, and a HC$_3$N is greatly enhanced.  Given the low number of comparison species, and the additional sources of uncertainty it is hard to draw strong conclusions, however it is clear that some chemical differences do exist between the two systems.}
 
{\cite{2006A&A...458..417M,2011A&A...535A.103M} present molecular line absorption spectra for a spiral galaxy at $z$=0.82 which is situated along a line of sight to quasar PKS 1830-211. These absorption measurements provide accurate measures of the abundance of many molecules in this system. The region of the spiral galaxy probed by these measurements is within a star-forming spiral arm about 2 kpc from the nucleus. The abundances of this object should hence reflect a less extreme, more normal star-forming environment than the others presented here. Various gas tracers are slightly suppressed in this object compared to Arp~157, including C$_2$H, HCN, HCO$^+$ and HNC, while methanol and H$_2$CO are suppressed further. Table \ref{goodfittable} shows that when taking into account the error bars the $\chi^2_r$ is very large. A quiescent star-forming disk is thus not likely a good model for the chemistry in Arp~157.}

NGC~1068 is one of the closest Seyfert galaxies, and the molecular material close to its AGN is pervaded by X-ray and cosmic ray radiation originating in the nuclear accretion disks \citep{1996ApJ...466..561M}. Its chemistry is thought to be dominated by a giant X-ray-dominated region (XDR) in its nucleus \citep[e.g.][]{2004A&A...419..897U}. 
When comparing NGC~1068 to Arp~157 in Figure \ref{abundfig}, it is apparent that Arp~157 has lower C$_2$H, HCN, HNC, C$^{18}$O, CN and HC$_3$N abundances.  These difference are also shown in Figure \ref{abundfig_diff}.
{The standard deviation for NGC~1068 in Table \ref{goodfittable} is large, however the $\chi^2$ (which takes into account the error bars) is low.
Arp~157 does not have a strong AGN, and so it is somewhat surprising that a prototype XDR can match reasonably well the observed abundances. Perhaps the violence of the ongoing merger enhances the hard radiation field in this object, simulating an XDR environment without a central AGN. }

NGC 253 is thought to be in an relatively early stage of starburst evolution, and has young stellar populations in its nucleus ($\approx$6 Myr; \citealt{2009MNRAS.392L..16F}). PDRs, although present \citep{2009ApJ...706.1323M}, do not characterise the ISM. Instead, the chemistry in the nucleus of NGC 253 is dominated by large-scale shocks \citep{2011A&A...525A..89A}. 
Of the galaxies present in Figure \ref{abundfig}, NGC~253 has the most similar molecular abundances to Arp~157 (see also Figure \ref{abundfig_diff}). This system has both the lowest standard deviation and the lowest $\chi^2_r$ of those studied here (see Table \ref{goodfittable}).

The main differences between Arp~157 and NGC~253 is that Arp~157 has a lower C$^{18}$O abundance (lower than any of the other galaxy studied), and a higher HCO$^+$ abundance. 
The HNCO molecule abundance in Arp~157 might also be marginally enhanced. Both HNCO and HCO$^+$ can be enhanced in shock dominated regions, or by an increased amount of incident X-ray radiation. As discussed above, perhaps Arp~157 has a hard radiation field like NGC~1068 (but caused by the merger), or perhaps stronger shocks could be present in Arp~157 than in NGC~253.  Unfortunately however (as Figure \ref{abundfig} shows) our observations are not deep enough to detect SiO emission, which is classically used as a `smoking gun' to determine if shocks are present. {In addition HNCO has been shown to be a good indicator of starburst stage \citep{2009ApJ...694..610M}, and so its enhanced abundance could indicate that Arp~157 has larger amounts of material fuelling its young starburst than NGC~253. }

\begin{table}
\caption{Goodness of fit statistics between other extragalactic objects and Arp~157. }
\begin{tabular*}{0.45\textwidth}{@{\extracolsep{\fill}}l l r r}
\hline
 Type & Name & Standard Deviation & $\chi^{2}_r$  \\
 (1) & (2) & (3) & (4) \\
 \hline
Galaxy & &\\
&   NGC1068&      0.57&      2.13\\
&    NGC253&      0.19&      2.02\\
&       M82&      0.38&     36.72\\
&    Arp220&      0.56&     33.63\\
&PKS1830-21&      0.56&  18125.29\\
     Galactic Regions & &\\
&   SgrB2(N)&      1.67&     14194\\
&  SgrB2(OH)&      0.12&       141\\
  &    TMC-1&      0.87&       264\\
   &   L134N&      0.72&       184\\
&  Orion Bar&      1.03&      4961\\
\hline
\end{tabular*}
\parbox[t]{0.45 \textwidth}{ \textit{Notes:} Column 1 contains the object type (Galactic region or galaxy). Column 2 contains the name of the region/object. Column 3 contains the standard deviation (in log space) of the residuals between the abundances of Arp~157 and the object, as presented in Figures \ref{abundfiggal_diff} and \ref{abundfig_diff}. Column 4 contains the reduced $\chi^2$ statistic which takes into account the error bars and number of molecules used in the comparison.}
\label{goodfittable}
\end{table}

\subsubsection{Comparison with Galactic regions}
\label{chem_abund_regions}

\begin{figure*}
\begin{center}
\includegraphics[width=0.95\textwidth,angle=0,clip,trim=0.0cm 0.0cm 0cm 0cm]{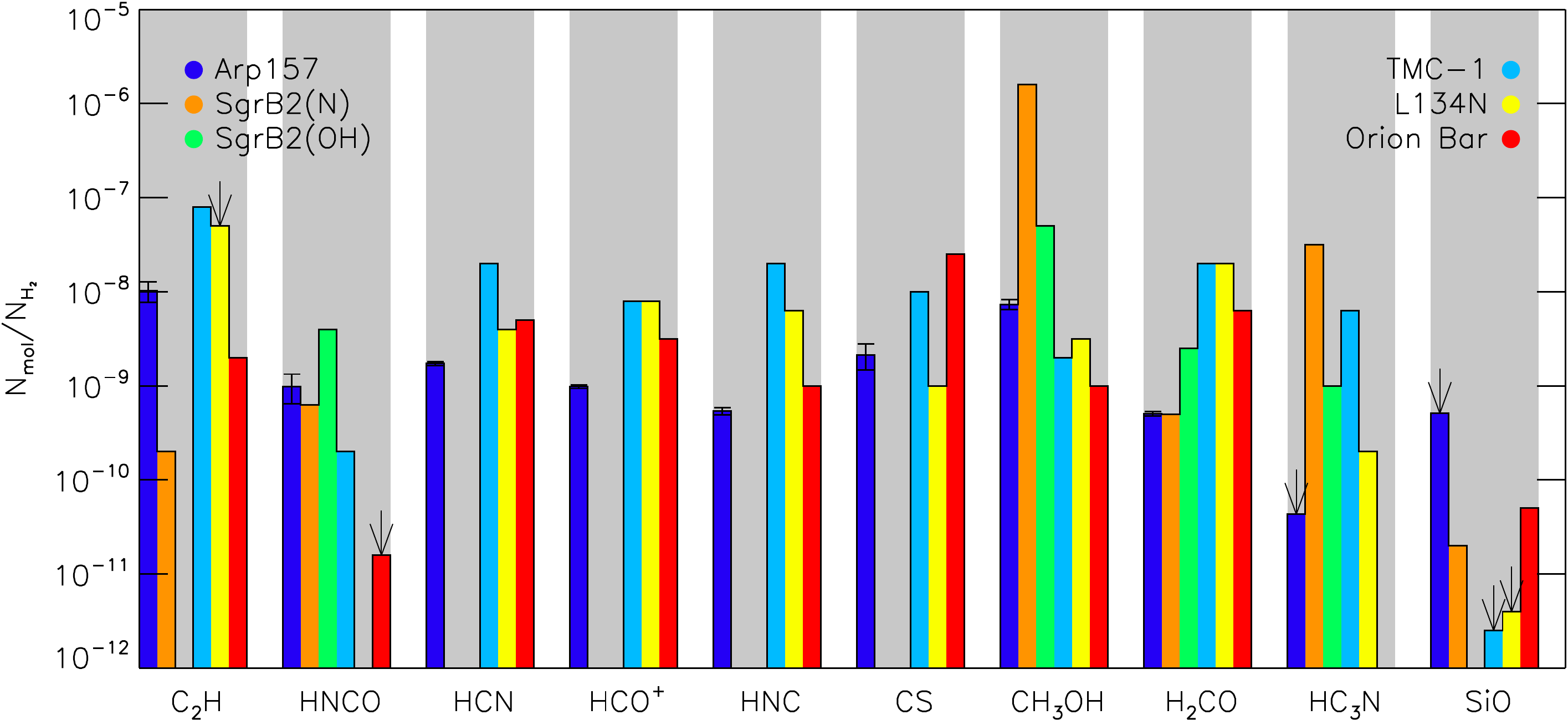}
 \end{center}
 \caption{Comparison of the fractional abundances in Arp~157 (blue), to the Galactic regions Sgr B2(N), Sgr B2(OH), TMC-1, L34N, and the Orion Bar. The Y-axis shows the total number of molecules of each species, normalised by H$_2$ (in the case of Arp~157 this has been estimated from the excitation analysis presented in Section \ref{cotex}). Grey shaded regions act as a guide to the eye for each set of molecules. Where no bar is shown, a measurement for this source is not included in \protect \cite{2006ApJS..164..450M}. Error bars indicate the error in the abundance ratio only for Arp~157. Arrows indicate upper limits.}
 \label{abundfiggal}
 \end{figure*}

\begin{figure*}
\begin{center}
\includegraphics[width=0.95\textwidth,angle=0,clip,trim=0.0cm 0.0cm 0cm 0cm]{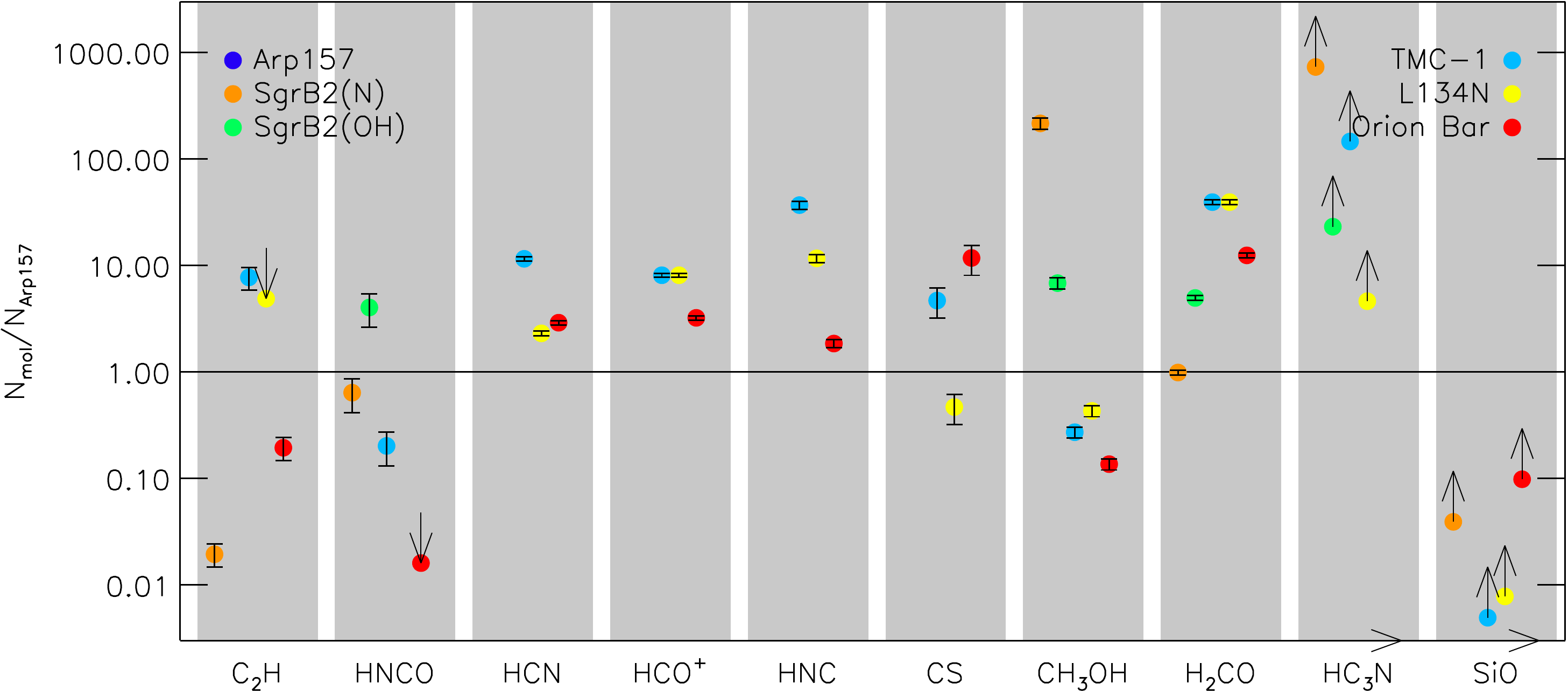}
 \end{center}
 \caption{Comparison of the molecular abundances (with respect to H$_2$) in the Galactic regions Sgr B2(N), Sgr B2(OH), TMC-1, L34N, and the Orion Bar, with respect to the abundance of that molecule in Arp~157 (with respect to H$_2$; blue). Molecules which were not detected in Arp~157 are included as upper limits, but all other details are as in Figure \protect \ref{abundfiggal}.}
 \label{abundfiggal_diff}
 \end{figure*}

 In Figure \ref{abundfiggal} we again show our derived abundances for the molecules detected in Arp~157. We also show the same values for a sample of Galactic molecular regions, whose abundance ratios were tabulated in \cite{2006ApJS..164..450M}.  These regions (Sgr B2(N), Sgr B2(OH), TMC-1, L34N, and the Orion Bar) are considered to be prototypes of their respective chemistry within the Galaxy. As CS abundances are not available for some regions, we show abundance ratios with respect to H$_2$. For the Galactic sources these come directly from \cite{2006ApJS..164..450M}. For Arp~157 we use the $^{13}$CO total column density derived from our excitation temperature analysis, and assume a Galactic H$_2$ to $^{13}$CO abundance ratio of 7$\times$10$^{5}$ (as before).  {Figure \ref{abundfiggal_diff} shows the molecular abundances of these sources, normalised by that found in this paper for Arp~157, as above. In several of the regions we do not have enough overlap to robustly determine which is most similar to Arp~157, and so we caution about drawing strong conclusions based on these values.}

The Orion Bar is included here as a prototype PDR, where the molecular emission originates from the surface of molecular clouds irradiated by far-ultraviolet photons from nearby young OB stars. As shown in Figure \ref{abundfiggal_diff}, many chemical species (e.g. C$_2$H, HNCO,  CH$_3$OH) are depressed in the Orion bar with respect to Arp~157, and H$_2$CO, CS, HCN and HCO$^+$ are enhanced. This region shows a high dispersion in Figure \ref{abundfiggal_diff}, and has a large standard deviation and $\chi^2_r$ in Table \ref{goodfittable}. Thus overall it does not seem that this Galactic PDR is a good template for the chemistry of Arp~157. 

TMC1 and L134N are quiescent cold dark clouds, whose molecular composition is dominated by gas-phase ion-molecule chemistry at sites of low-mass star formation. These sources are chemically quite similar to Arp~157 in many of the more common gas tracers, with similar abundances of CS, HCN, and CH$_3$OH. However, these quiescent cold dark clouds are over-abundant in C$_2$H, HCO$^+$, HNC and H$_2$CO, and under-abundant in HNCO. L134N shows more similarities to Arp~157 then TMC1.

Sgr B2 is a giant molecular cloud in the central region of our galaxy. The northern region, Sgr B2(N), is believed to be dominated by hot core chemistry, associated with massive star formation near the nucleus of the Milky Way. Its chemistry leads to an enhanced CH$_3$OH and HC$_3$N abundance with respect to Arp~157.  Sgr B2(N) also shows a decreased abundance of C$_2$H with respect to Arp~157 (see Figure \ref{abundfiggal_diff}). We note that if the questionable detection of SO$_2$ discussed in Section \ref{linesfound} is real, this would be the only source we consider which has a higher abundance of SO$_2$ than Arp~157. The large standard deviation value in Table \ref{goodfittable} suggests that a region dominated by hot core emission does not provide a good template for the chemistry of Arp~157.

Sgr B2(OH) is another part of the larger Sgr B2 galactic centre molecular cloud. This position within the cloud has little emission from hot cores \citep{2006A&A...455..971R}, and instead shock desorption of ices is thought to be an important driver of the gas phase chemistry \citep{1997ApJ...482L..45M}. This region most resembles Arp~157 in Figure \ref{abundfiggal_diff}, and has the lowest standard deviation in Table \ref{goodfittable}. {However, we caution that this conclusion is based on comparison of only 4 molecules, and should be treated with caution}. The suggested similarity between this source and Arp~157 is in some ways unsurprising, as it also has been found to have similar chemistry to NGC~253 \citep{2006ApJS..164..450M}, with similar (but slightly lower) HNCO, CH$_3$OH, H$_2$CO and HC$_3$N abundances. 

\section{Conclusions}
\label{conclude}

 In this work, we investigated the physical conditions within the molecular gas reservoir of interacting LIRG Arp~157 (NGC520).
 We presented an unbiased survey searching for molecular emission in this galaxy at $\lambda$=3mm (and supporting data from the $\lambda$=2mm and 1mm atmospheric windows).
 We detect 19 different lines, from 12 different molecular species. The lines are asymmetric, with profiles that differ systematically as a function of critical density. We interpret this as observing an edge-on, asymmetric starburst ring, plus a central molecular component, which is less dense, and more optically thin.

We derive the physical conditions within the gas reservoir of this system using rotation diagrams, and an excitation analysis. 
Analysis of isotope ratios and the excitation of CO suggests that the bulk of the molecular gas in Arp~157 is reasonably cool, with temperatures of $\approx$15-30K.
The temperatures we derive from our rotation diagrams vary little between species, and are lower than our estimates of the excitation temperatures. This may suggest that the high $J$ emitting gas is more concentrated than the low excitation gas in this system.

We compare our molecular abundances with several well studied galaxies, and with Galactic regions, in order to determine the primary driver of the chemistry in Arp~157. Of the nearby galaxies we considered, NGC~253 has the most similar molecular abundances to Arp~157. NGC 253 is thought to be in an early stage of starburst evolution, and the chemistry in its nucleus is dominated by large-scale shocks. {This suggestion that Arp~157 could have a shock-dominated ISM is tentatively supported by a comparison with galactic regions, where Sgr B2(OH) (where shock desorption of grains is thought to be an important driver of the gas phase chemistry) has the most similar abundances to Arp~157}.  {The XDR in the centre of NGC~1068 also shows similar abundance patterns to those found in Arp~157, and we postulate that perhaps the violence of the ongoing merger enhances the hard radiation field in this object, simulating an XDR environment without a central AGN. } 
Overall we find that shocks and/or X-rays may be important drivers of the chemistry in Arp~157. If these are caused by the ongoing merger or a starburst wind will be investigated further in a future work.

{We also compared this galaxy with the ULIRG Arp~220. Although the number of species available for comparison is low, it seems that Arp~157 has significant chemical differences to this more extreme starburst. 
This suggests that different physical mechanisms may become important in driving gas chemistry at different points in the evolution of a merger induced starburst. In addition it raises the possibility of calibrating chemical diagnostics which can determine how advanced, and how extreme conditions are within star-bursting merger remnants.}

In order to further determine the configuration of the gas in the system, and what drives the ISM chemistry, high resolution (sub-arcsecond) interferometric observations of different molecular species will be required. Comparison of dense gas tracers, which we postulate may lie in a starburst ring, and optically thin CO isotopologues which are abundant in the kinematic centre of the system will enable us to gain a better understanding of the configuration of the gas, and understand the origin of the asymmetric emission in this system. Detections of pure shock tracers, such as SiO would also help confirm if the ISM is dominated by shocks, or if X-rays are a more important chemical driver. 
Clearly similar line surveys of other (U)LIRGs will be required in order to determine if this galaxy is special, and if chemical diagnostics can be developed that trace the changing conditions along the merger sequence.

 \vspace{0.5cm}
\noindent \textbf{Acknowledgments}

The authors would like to thank Izaskun Jim\'enez-Serra, Sergio Mart\'in, Estelle Bayet and Roberto Galv\'an Madrid for help and advice during the preparation of this manuscript.
The research leading to these results has received funding from the European
Community's Seventh Framework Programme (/FP7/2007-2013/) under grant agreement
No 229517. A.H., and  N.J.E. acknowledge support for this
work from NSF Grant AST--1109116 to the University of Texas at Austin, and the State of
Texas. This paper is based on observations carried out with the IRAM Thirty Meter Telescope. IRAM is supported by INSU/CNRS (France), MPG (Germany) and IGN (Spain). This research has made use of the NASA/IPAC Extragalactic Database (NED) which is operated by the Jet Propulsion Laboratory, California Institute of Technology, under contract with the National Aeronautics and Space Administration.

\bsp
\bibliographystyle{mn2e}
\bibliography{bibArp157}

\begin{thebibliography}{47}
\expandafter\ifx\csname natexlab\endcsname\relax\def\natexlab#1{#1}\fi

\bibitem[{Aladro {et~al}\mbox{.}(2011{\natexlab{a}})Aladro, Mart{\'\i}n,
  Mart{\'\i}n-Pintado, Mauersberger, Henkel, Oca{\~n}a~Flaquer, \&
  Amo-Baladr{\'o}n}]{2011A&A...535A..84A}
Aladro R., Mart{\'\i}n S., Mart{\'\i}n-Pintado J., Mauersberger R., Henkel C.,
  Oca{\~n}a~Flaquer B., Amo-Baladr{\'o}n M.~A., 2011{\natexlab{a}}, Astronomy
  and Astrophysics, 535, 84

\bibitem[{Aladro {et~al}\mbox{.}(2011{\natexlab{b}})Aladro,
  Mart{\'\i}n-Pintado, Mart{\'\i}n, Mauersberger, \&
  Bayet}]{2011A&A...525A..89A}
Aladro R., Mart{\'\i}n-Pintado J., Mart{\'\i}n S., Mauersberger R., Bayet E.,
  2011{\natexlab{b}}, Astronomy and Astrophysics, 525, 89

\bibitem[{Aladro {et~al}\mbox{.}(2013)Aladro, Viti, Bayet, Riquelme,
  Mart{\'\i}n, Mauersberger, Mart{\'\i}n-Pintado, Requena-Torres, Kramer, \&
  Wei{\ss}}]{2013A&A...549A..39A}
Aladro R. {et~al.}, 2013, Astronomy and Astrophysics, 549, 39

\bibitem[{Arp \& Madore(1987)}]{1987QB857.A76......}
Arp H.~C., Madore B., 1987, Cambridge ; New York : Cambridge University Press,
  76

\bibitem[{Bayet {et~al}\mbox{.}(2009{\natexlab{a}})Bayet, Aladro, Mart{\'\i}n,
  Viti, \& Mart{\'\i}n-Pintado}]{2009ApJ...707..126B}
Bayet E., Aladro R., Mart{\'\i}n S., Viti S., Mart{\'\i}n-Pintado J.,
  2009{\natexlab{a}}, The Astrophysical Journal, 707, 126

\bibitem[{Bayet {et~al}\mbox{.}(2012)Bayet, Davis, Bell, \&
  Viti}]{2012MNRAS.424.2646B}
Bayet E., Davis T.~A., Bell T.~A., Viti S., 2012, Monthly Notices of the Royal
  Astronomical Society, 424, 2646

\bibitem[{Bayet {et~al}\mbox{.}(2009{\natexlab{b}})Bayet, Viti, Williams,
  Rawlings, \& Bell}]{2009ApJ...696.1466B}
Bayet E., Viti S., Williams D.~A., Rawlings J. M.~C., Bell T.,
  2009{\natexlab{b}}, The Astrophysical Journal, 696, 1466

\bibitem[{Beswick {et~al}\mbox{.}(2003)Beswick, Pedlar, Clemens, \&
  Alexander}]{2003MNRAS.346..424B}
Beswick R.~J., Pedlar A., Clemens M.~S., Alexander P., 2003, Monthly Notices of
  the Royal Astronomical Society, 346, 424

\bibitem[{Crocker {et~al}\mbox{.}(2012)Crocker, Krips, Bureau, Young, Davis,
  Bayet, Alatalo, Blitz, Bois, Bournaud, Cappellari, Davies, de~Zeeuw, Duc,
  Emsellem, Khochfar, Krajnovic, Kuntschner, Lablanche, McDermid, Morganti,
  Naab, Oosterloo, Sarzi, Scott, Serra, \& Weijmans}]{2012MNRAS.421.1298C}
Crocker A. {et~al.}, 2012, Monthly Notices of the Royal Astronomical Society,
  421, 1298

\bibitem[{Davis {et~al}\mbox{.}(2013)Davis, Bayet, Crocker, Topal, \&
  Bureau}]{2013MNRAS.433.1659D}
Davis T.~A., Bayet E., Crocker A., Topal S., Bureau M., 2013, Monthly Notices
  of the Royal Astronomical Society, 433, 1659

\bibitem[{Fern{\'a}ndez-Ontiveros, Prieto \&
  Acosta-Pulido(2009)Fern{\'a}ndez-Ontiveros, Prieto, \&
  Acosta-Pulido}]{2009MNRAS.392L..16F}
Fern{\'a}ndez-Ontiveros J.~A., Prieto M.~A., Acosta-Pulido J.~A., 2009, Monthly
  Notices of the Royal Astronomical Society: Letters, 392, L16

\bibitem[{F{\"o}rster~Schreiber {et~al}\mbox{.}(2009)F{\"o}rster~Schreiber,
  Genzel, Bouche, Cresci, Davies, Buschkamp, Shapiro, Tacconi, Hicks, Genel,
  Shapley, Erb, Steidel, Lutz, Eisenhauer, Gillessen, Sternberg, Renzini,
  Cimatti, Daddi, Kurk, Lilly, Kong, Lehnert, Nesvadba, Verma, McCracken,
  Arimoto, Mignoli, \& Onodera}]{2009ApJ...706.1364F}
F{\"o}rster~Schreiber N.~M. {et~al.}, 2009, The Astrophysical Journal, 706,
  1364

\bibitem[{Garc{\'\i}a-Burillo {et~al}\mbox{.}(2010)Garc{\'\i}a-Burillo, Usero,
  Fuente, Mart{\'\i}n-Pintado, Boone, Aalto, Krips, Neri, Schinnerer, \&
  Tacconi}]{2010A&A...519A...2G}
Garc{\'\i}a-Burillo S. {et~al.}, 2010, Astronomy and Astrophysics, 519, 2

\bibitem[{Genzel {et~al}\mbox{.}(2012)Genzel, Tacconi, Combes, Bolatto, Neri,
  Sternberg, Cooper, Bouche, Bournaud, Burkert, Comerford, Cox, Davis,
  F{\"o}rster~Schreiber, Garc{\'\i}a-Burillo, Graci{\'a}-Carpio, Lutz, Naab,
  Newman, Saintonge, Shapiro, Shapley, \& Weiner}]{2012ApJ...746...69G}
Genzel R. {et~al.}, 2012, The Astrophysical Journal, 746, 69

\bibitem[{Glassgold, Galli \& Padovani(2012)Glassgold, Galli, \&
  Padovani}]{2012ApJ...756..157G}
Glassgold A.~E., Galli D., Padovani M., 2012, The Astrophysical Journal, 756,
  157

\bibitem[{Goldsmith \& Langer(1999)}]{1999ApJ...517..209G}
Goldsmith P.~F., Langer W.~D., 1999, The Astrophysical Journal, 517, 209

\bibitem[{Heiderman {et~al}\mbox{.}(2011)Heiderman, Evans, Gebhardt, Blanc,
  Davis, Papovich, Iono, \& Yun}]{2011nha..confE..29H}
Heiderman A.~L., Evans N. J.~I., Gebhardt K., Blanc G., Davis T.~A., Papovich
  C., Iono D., Yun M.~S., 2011, New Horizons in Astronomy, 29

\bibitem[{Henkel \& Mauersberger(1993)}]{1993A&A...274..730H}
Henkel C., Mauersberger R., 1993, Astronomy and Astrophysics, 274, 730

\bibitem[{Huettemeister {et~al}\mbox{.}(1995)Huettemeister, Henkel,
  Mauersberger, Brouillet, Wiklind, \& Millar}]{1995A&A...295..571H}
Huettemeister S., Henkel C., Mauersberger R., Brouillet N., Wiklind T., Millar
  T.~J., 1995, Astronomy and Astrophysics, 295, 571

\bibitem[{Konstantopoulos {et~al}\mbox{.}(2009)Konstantopoulos, Bastian, Smith,
  Westmoquette, Trancho, \& Gallagher}]{2009ApJ...701.1015K}
Konstantopoulos I.~S., Bastian N., Smith L.~J., Westmoquette M.~S., Trancho G.,
  Gallagher J. S.~I., 2009, The Astrophysical Journal, 701, 1015

\bibitem[{Kotilainen {et~al}\mbox{.}(2001)Kotilainen, Reunanen, Laine, \&
  Ryder}]{2001A&A...366..439K}
Kotilainen J.~K., Reunanen J., Laine S., Ryder S.~D., 2001, Astronomy and
  Astrophysics, 366, 439

\bibitem[{Maloney, Hollenbach \& Tielens(1996)Maloney, Hollenbach, \&
  Tielens}]{1996ApJ...466..561M}
Maloney P.~R., Hollenbach D.~J., Tielens A. G. G.~M., 1996, Astrophysical
  Journal v.466, 466, 561

\bibitem[{Mao {et~al}\mbox{.}(2000)Mao, Henkel, Schulz, Zielinsky,
  Mauersberger, St{\"o}rzer, Wilson, \& Gensheimer}]{2000A&A...358..433M}
Mao R.~Q., Henkel C., Schulz A., Zielinsky M., Mauersberger R., St{\"o}rzer H.,
  Wilson T.~L., Gensheimer P., 2000, Astronomy and Astrophysics, 358, 433

\bibitem[{Mart{\'\i}n {et~al}\mbox{.}(2010)Mart{\'\i}n, Aladro,
  Mart{\'\i}n-Pintado, \& Mauersberger}]{2010A&A...522A..62M}
Mart{\'\i}n S., Aladro R., Mart{\'\i}n-Pintado J., Mauersberger R., 2010,
  Astronomy and Astrophysics, 522, 62

\bibitem[{Mart{\'\i}n {et~al}\mbox{.}(2011)Mart{\'\i}n, Krips,
  Mart{\'\i}n-Pintado, Aalto, Zhao, Peck, Petitpas, Monje, Greve, \&
  An}]{2011A&A...527A..36M}
Mart{\'\i}n S. {et~al.}, 2011, Astronomy and Astrophysics, 527, 36

\bibitem[{Martin, Mart{\'\i}n-Pintado \& Mauersberger(2009)Martin,
  Mart{\'\i}n-Pintado, \& Mauersberger}]{2009ApJ...694..610M}
Martin S., Mart{\'\i}n-Pintado J., Mauersberger R., 2009, The Astrophysical
  Journal, 694, 610

\bibitem[{Mart{\'\i}n {et~al}\mbox{.}(2005)Mart{\'\i}n, Mart{\'\i}n-Pintado,
  Mauersberger, Henkel, \& Garc{\'\i}a-Burillo}]{2005ApJ...620..210M}
Mart{\'\i}n S., Mart{\'\i}n-Pintado J., Mauersberger R., Henkel C.,
  Garc{\'\i}a-Burillo S., 2005, The Astrophysical Journal, 620, 210

\bibitem[{Mart{\'\i}n, Mart{\'\i}n-Pintado \& Viti(2009)Mart{\'\i}n,
  Mart{\'\i}n-Pintado, \& Viti}]{2009ApJ...706.1323M}
Mart{\'\i}n S., Mart{\'\i}n-Pintado J., Viti S., 2009, The Astrophysical
  Journal, 706, 1323

\bibitem[{Mart{\'\i}n {et~al}\mbox{.}(2006)Mart{\'\i}n, Mauersberger,
  Mart{\'\i}n-Pintado, Henkel, \& Garc{\'\i}a-Burillo}]{2006ApJS..164..450M}
Mart{\'\i}n S., Mauersberger R., Mart{\'\i}n-Pintado J., Henkel C.,
  Garc{\'\i}a-Burillo S., 2006, The Astrophysical Journal Supplement Series,
  164, 450

\bibitem[{Mart{\'\i}n-Pintado {et~al}\mbox{.}(1997)Mart{\'\i}n-Pintado,
  de~Vicente, Fuente, \& Planesas}]{1997ApJ...482L..45M}
Mart{\'\i}n-Pintado J., de~Vicente P., Fuente A., Planesas P., 1997,
  Astrophysical Journal Letters v.482, 482, L45

\bibitem[{Mauersberger \& Henkel(1989)}]{1989A&A...223...79M}
Mauersberger R., Henkel C., 1989, Astronomy and Astrophysics, 223, 79

\bibitem[{M{\"u}ller {et~al}\mbox{.}(2005)M{\"u}ller, Schl{\"o}der, Stutzki, \&
  Winnewisser}]{2005JMoSt.742..215M}
M{\"u}ller H. S.~P., Schl{\"o}der F., Stutzki J., Winnewisser G., 2005, Journal
  of Molecular Structure, 742, 215

\bibitem[{M{\"u}ller {et~al}\mbox{.}(2001)M{\"u}ller, Thorwirth, Roth, \&
  Winnewisser}]{2001A&A...370L..49M}
M{\"u}ller H. S.~P., Thorwirth S., Roth D.~A., Winnewisser G., 2001, Astronomy
  and Astrophysics, 370, L49

\bibitem[{Muller {et~al}\mbox{.}(2011)Muller, Beelen, Guelin, Aalto, Black,
  Combes, Curran, Theule, \& Longmore}]{2011A&A...535A.103M}
Muller S. {et~al.}, 2011, Astronomy and Astrophysics, 535, 103

\bibitem[{Muller {et~al}\mbox{.}(2006)Muller, Guelin, Dumke, Lucas, \&
  Combes}]{2006A&A...458..417M}
Muller S., Guelin M., Dumke M., Lucas R., Combes F., 2006, Astronomy and
  Astrophysics, 458, 417

\bibitem[{Paggi {et~al}\mbox{.}(2013)Paggi, Fabbiano, Risaliti, Wang, \&
  Elvis}]{2013arXiv1303.2630P}
Paggi A., Fabbiano G., Risaliti G., Wang J., Elvis M., 2013, arXiv, 2630

\bibitem[{Papadopoulos(2007)}]{2007ApJ...656..792P}
Papadopoulos P.~P., 2007, The Astrophysical Journal, 656, 792

\bibitem[{Requena-Torres {et~al}\mbox{.}(2006)Requena-Torres,
  Mart{\'\i}n-Pintado, Rodr{\'\i}guez-Franco, Mart{\'\i}n,
  Rodr{\'\i}guez-Fern{\'a}ndez, \& de~Vicente}]{2006A&A...455..971R}
Requena-Torres M.~A., Mart{\'\i}n-Pintado J., Rodr{\'\i}guez-Franco A.,
  Mart{\'\i}n S., Rodr{\'\i}guez-Fern{\'a}ndez N.~J., de~Vicente P., 2006,
  Astronomy and Astrophysics, 455, 971

\bibitem[{Solomon, Downes \& Radford(1992)Solomon, Downes, \&
  Radford}]{1992ApJ...387L..55S}
Solomon P.~M., Downes D., Radford S. J.~E., 1992, Astrophysical Journal, 387,
  L55

\bibitem[{Stanford(1991)}]{1991ApJ...381..409S}
Stanford S.~A., 1991, Astrophysical Journal, 381, 409

\bibitem[{Stanford \& Balcells(1991)}]{1991ApJ...370..118S}
Stanford S.~A., Balcells M., 1991, Astrophysical Journal, 370, 118

\bibitem[{Tully(1988)}]{1988ngc..book.....T}
Tully R.~B., 1988, Cambridge and New York

\bibitem[{Usero {et~al}\mbox{.}(2004)Usero, Garc{\'\i}a-Burillo, Fuente,
  Mart{\'\i}n-Pintado, \& Rodr{\'\i}guez-Fern{\'a}ndez}]{2004A&A...419..897U}
Usero A., Garc{\'\i}a-Burillo S., Fuente A., Mart{\'\i}n-Pintado J.,
  Rodr{\'\i}guez-Fern{\'a}ndez N.~J., 2004, Astronomy and Astrophysics, 419,
  897

\bibitem[{van Dishoeck \& Black(1988)}]{1988ApJ...334..771V}
van Dishoeck E.~F., Black J.~H., 1988, Astrophysical Journal, 334, 771

\bibitem[{van Dishoeck {et~al}\mbox{.}(1992)van Dishoeck, Glassgold, Guelin,
  Jaffe, Neufeld, Tielens, \& Walmsley}]{1992IAUS..150..285V}
van Dishoeck E.~F., Glassgold A.~E., Guelin M., Jaffe D.~T., Neufeld D.~A.,
  Tielens A. G. G.~M., Walmsley C.~M., 1992, in Astrochemistry of Cosmic
  Phenomena: Proceedings of the 150th Symposium of the International
  Astronomical Union, p. 285

\bibitem[{Wilson \& Rood(1994)}]{1994ARA&A..32..191W}
Wilson T.~L., Rood R., 1994, Annual Review of Astronomy and Astrophysics, 32,
  191

\bibitem[{Yun \& Hibbard(2001)}]{2001ApJ...550..104Y}
Yun M.~S., Hibbard J.~E., 2001, The Astrophysical Journal, 550, 104

\end{thebibliography}
\bibdata{bibArp157}
\bibstyle{mn2e}

\label{lastpage}

\end{document}